\documentclass[a4paper,fleqn,usenatbib]{mnras}

\usepackage{newtxtext,newtxmath}
\usepackage[T1]{fontenc}
\usepackage{ae,aecompl}

\usepackage{graphicx}    % Including figure files
\usepackage{amsmath}    % Advanced maths commands
\usepackage{amssymb}    % Extra maths symbols
\usepackage{subcaption}
\usepackage{bm}

\newcommand{\kms}{km\,s$^{-1}$}
\newcommand{\mpch}{\,$h^{-1}$\,Mpc}
\newcommand{\hmpc}{\,$h$\,Mpc$^{-1}$}

\newcommand{\msol}{$h^{-1}$\,M$_{\sun}$}
\newcommand{\hubbleunit}{$h$\,km\,s$^{-1}$\,Mpc$^{-1}$}
\newcommand{\angstrom}{\mbox{\normalfont\AA}}

\title[Density-velocity cross-correlation in 6dFGS]{Joint growth rate measurements from redshift-space distortions and peculiar velocities in the 6dF Galaxy Survey}

\author[Adams and Blake]{
	Caitlin Adams$^{1}$
	and Chris Blake$^{1}$\thanks{E-mail: cblake@swin.edu.au}
	\\
	$^{1}$Centre for Astrophysics \& Supercomputing, Swinburne University of Technology, P.O. Box 218, Hawthorn, VIC 3122, Australia.
}

\date{Accepted XXX. Received YYY; in original form ZZZ}

\pubyear{2020}

\begin{document}
\label{firstpage}
\pagerange{\pageref{firstpage}--\pageref{lastpage}}
\maketitle

\begin{abstract}
We present a new model for the cross-covariance between galaxy redshift-space distortions and peculiar velocities. 
We combine this with the auto-covariance models of both probes in a fully self-consistent, maximum-likelihood method, allowing us to extract enhanced cosmological parameter constraints. 
When applying our method to the 6-degree Field Galaxy Survey (6dFGS), our constraint on the growth rate of structure is $f\sigma_8 = 0.384 \pm 0.052 \rm{(stat)} \pm 0.061 \rm{(sys)}$ and our constraint for the redshift-space distortion parameter is $\beta = 0.289^{+0.044}_{-0.043} \rm{(stat)} \pm 0.049 \rm{(sys)}$.
We find that the statistical uncertainty for the growth rate of structure is reduced by 64\% when using the complete covariance model compared to the redshift-space distortion auto-covariance model and 50\% when compared to using the peculiar velocity auto-covariance model. 
Our constraints are consistent with those from the literature on combining multiple tracers of large-scale structure, as well as those from other 6dFGS analyses. 
Our measurement is also consistent with the standard cosmological model. 
\end{abstract}


%%%%%%%%%%%%%%%%%%%%%%%%%%%%%%%%%%%%%%%%%%%%%%%%%%

%%%%%%%%%%%%%%%%% BODY OF PAPER %%%%%%%%%%%%%%%%%%

\section{Introduction}\label{sec:introduction}
%This is a simple template for authors to write new MNRAS papers.
%See \texttt{mnras\_sample.tex} for a more complex example, and \texttt{mnras\_guide.tex}
%for a full user guide.
The current cosmological model explains the observed accelerating expansion of the Universe by including a dark energy component in Einstein's general relativistic field equations.
While this model is supported by numerous high- and low-redshift observations, such as the cosmic microwave background \citep[CMB; e.g.][]{PlanckCollaboration2018}, baryon acoustic oscillations \citep[BAO; e.g.][]{Ata2017} and type Ia supernovae \citep[SNe Ia; e.g.][]{Scolnic2017}, we have a poor understanding of the physical nature of dark energy.
A possible alternative theory for the accelerating expansion would be that gravity behaves differently on cosmological scales, generally classed as modified gravity theories. 
However, we require higher precision tests on large scales if we are to distinguish between dark energy and modified gravity.

Modifying gravity directly affects how structures like galaxies and galaxy clusters form in the Universe.
Therefore, one of the key ways we can test gravity on large scales is through the linear growth rate of structure, $f(a) \equiv d\ln(D(a))/d\ln(a)$, where $D$ is the linear growth function describing the growth of matter perturbations and $a$ is the scale factor, which characterises the Universe's expansion.
General relativity predicts that the value of the growth rate should be a function of the total matter density through $f(a) \approx \Omega_m(a)^{0.55}$ \citep{Linder2005}.
Precise measurements of the growth rate may help to rule out dark energy or alternative models.

The growth rate of structure can be constrained by two key cosmological probes: direct measurements of peculiar velocities and statistical redshift space distortions.
A galaxy's peculiar velocity is its motion due to gravitational attraction, rather than its motion due to cosmological expansion. 
A galaxy's total motion can be inferred from its redshift, and the peculiar velocity contribution may be extracted if one has a redshift-independent distance measure of the galaxy's position, such as through the Fundamental Plane or Tully-Fisher methods.
Importantly, peculiar velocities are directly proportional to the underlying matter overdensity field via the growth rate, and at linear scales, peculiar velocities are considered to be unbiased tracers of the matter overdensity field \citep{Desjacques2010}.
Because of their contribution to a galaxy's total redshift, peculiar velocities are also statistically encoded in the distribution of large-scale structure in redshift space, a signal known as the redshift-space distortion (RSD). 
Originally modelled by \cite{Kaiser1987}, redshift-space distortions allow us to also constrain the growth rate of structure without requiring a redshift-independent distance estimator. 
RSD studies have constrained the growth rate at the level of 10\% at low redshifts ($z < 2$) \citep[e.g.][]{Blake2011b, Beutler2012, Alam2016, Ruggeri2018}.

Peculiar velocities and redshift-space distortions constrain the growth rate on different physical scales; peculiar velocities are sensitive to the underlying matter overdensity field on scales of hundreds of \mpch, where redshift-space distortions are sensitive on scales of tens of \mpch.
This makes them highly-complementary when it comes to constraining the growth rate.
One way to take advantage of this complementarity is through density-velocity comparison methods, which use gravitational instability theory to predict one field from the other, then constrain the ratio of the growth rate and galaxy bias $\beta\equiv f/b$ by comparing the predicted and observed fields \citep[e.g.][]{Pike2005, Davis2011,Carrick2015}.
Multiple works have also shown that combining correlated tracers of the matter overdensity field leads to improved constraints on cosmological parameters \citep[e.g.][]{McDonald2009, Gil-Marin2010, Bernstein2011, Abramo2013, Blake2013}.
\cite{Adams2017} demonstrated that this effect extends to combining peculiar velocities and galaxy overdensities; by utilising the cross-correlation of these probes, they were able to improve the constraint on the growth rate by $\sim20\%$ compared to treating the probes independently.

This study builds upon two existing works: that of \cite{Johnson2014}, who constrained the growth rate of structure by modelling the peculiar velocity auto-correlation; and that of \cite{Adams2017}, who constrained the growth rate by modelling the cross-correlation between the galaxy overdensity and peculiar velocity fields, in addition to their auto-correlations.
We note that \cite{Adams2017} did not include a complete model of redshift-space distortions, meaning the galaxy overdensity field was only a function of the linear galaxy bias.
In this study, we aim to update the model to include a fully self-consistent description of redshift-space distortions, meaning that the galaxy overdensity field is a function of the linear galaxy bias and the growth rate of structure. 
We hypothesise that this will provide tighter constraints on the growth rate of structure, due to the complementary information provided by redshift-space distortions and peculiar velocities.
We validate our model by applying it to mock catalogues that are significantly more sophisticated than those utilised by \cite{Adams2017}; we then apply our analysis to data from the 6-degree Field Galaxy Survey (6dFGS). 
We also extend the work by \cite{Adams2017} by conducting a thorough analysis of the model systematics, including the calculation of a systematic uncertainty contribution to our growth rate of structure constraint.

We begin by introducing the data and simulations in Section \ref{sec:data_sims}.
We discuss the theory and methodology in Section \ref{sec:theory_methodology}.
The results from our tests on simulations are given in Section \ref{sec:simresults}, followed by those from the 6dFGS data in Section \ref{sec:6dfgsresults}.
Finally, we conclude with a summary in Section \ref{sec:summary}.

\section{Data and Simulations}\label{sec:data_sims}
We use measurements of the galaxy overdensity and peculiar velocity field from galaxy redshift surveys to constrain the growth rate of structure.
In this section, we cover the observed and simulated data used in our analysis.

\subsection{6dFGS} \label{subsec:data}
We utilise data from the 6-degree Field Galaxy Survey \citep[6dFGS;][]{Jones2005,Jones2006,Jones2009}.
Conducted on the UK Schmidt Telescope, the survey covers the southern sky (excluding 10 degrees around the galactic plane) out to redshift $z\approx0.23$.
The survey consists of two key samples: the redshift sample, 6dFGSz, and the peculiar velocity sample, 6dFGSv; we work with both in this analysis.

As of the survey's completion, 6dFGSz contains 125,071 redshifts, with a median redshift of $z=0.053$.
We draw our galaxy redshift sample from that used in the 6dFGS baryon acoustic oscillation analysis by \cite{Beutler2011}, which selected galaxies from 6dFGSz that had magnitude $K \leq 12.9$ in sky regions with greater than 60\% completeness, producing a total of 75,117 galaxies.
We impose an additional selection in limiting the redshift range to be $z \leq 0.1$, leaving us with 70,467 galaxies.
This restriction allows us to extract the redshift-space distortion signal in the data while balancing the computational requirements of our method, which scale with sample volume. 

It is important to note that our galaxy overdensity sample is not volume-limited, which means the galaxy bias of the sample will evolve with redshift.
As in previous redshift-space distortion studies \citep[e.g.][]{Beutler2012}, we may conveniently model the sample as having a single effective bias over the whole redshift range.
We note that this has some consequences for our modelling of the cross-correlation between the peculiar velocity and galaxy overdensity field, which may sample a different effective galaxy bias to the galaxy overdensity field; we discuss the evidence and proposed model for this effect in Section \ref{subsubsec:alphab}.

The final 6dFGSv sample consists of 9794 Fundamental Plane measurements in the redshift range $z\leq0.057$, with distance errors of around 26\% \citep{Magoulas2012}.
We use the sample as defined by \cite{Springob2014}, who selected galaxies with signal-to-noise ratios of greater than 5\angstrom$^{-1}$, and velocity dispersions greater than the resolution limit of the 6dF spectrograph ($\sigma_0 \geq$112\kms). 
This selection yielded logarithmic distance ratios for 8,885 galaxies, where the logarithmic distance ratio $\eta$ is related to the peculiar velocity $v_p$, but is preferred because of its Gaussian distribution \citep[see][]{Johnson2014}.

As in \cite{Adams2017}, we grid our galaxy redshift and logarithmic distance ratio samples, which allows us to simultaneously smooth over non-linear effects and reduce the computational requirements of our analysis (see Section \ref{subsubsection:gridding}).
We used a gridding scale of $L_{\delta} = 30$\mpch \ for the galaxy redshift sample, and a gridding scale of $L_{\eta} = 20$\mpch \ for the logarithmic distance ratio sample.
This resulted in $N_{\delta} = 1633$ grid cells for our redshift sample, and $N_{\eta} = 908$ grid cells for our logarithmic distance ratio sample.

Gridding the redshift sample allows us to directly calculate the galaxy overdensity in each cell by comparing the number of galaxies in the cell $N_{\text{cell}}$ to the number expected for the cell $N_{\text{exp}}$ through $\delta_g = \frac{N_{\text{cell}}}{N_{\text{exp}}} - 1$. 
We estimated $N_{\text{exp}}$ from the survey selection function, which we generated by combining the survey luminosity function with the position-dependent magnitude completeness \citep{Jones2006}.
These values were then normalised so that the total $N_{\text{exp}}$ value matched the total number of galaxies in our redshift sample and used to calculate the overdensity.
The shot noise for each galaxy overdensity measurement is determined from Poisson statistics as $\sigma_{\delta_g} = 1/\sqrt{N_{\text{exp}}}$.

We calculate the logarithmic distance ratio measurement for each cell as the average of all measurements within that cell.
The observational uncertainties are added in quadrature, giving an error in the mean. 
We refer the reader to \cite{Abate2008} for the motivation behind this approach, and note that we discuss how it affects the modelling of the logarithmic distance ratio auto-covariance in Section \ref{subsubsection:gridding}.

\subsection{Simulations}\label{subsec:sims}
We use the data and random mock catalogues developed by \cite{Carter2018}, which include accurate modelling of the halo occupation distribution of 6dFGS. 
Each mock is drawn from a unique COmoving Lagrangian Acceleration \citep[COLA;][]{Tassev2013} simulation, where each simulation has a box-length of $1.2 \ h^{-1}$Gpc and contains $(1728)^3$ particles with a mass resolution of $2.8 \times 10^{10}$\msol.
The advantage in using the COLA method is that it is faster than standard N-body techniques; the speed-gain comes from sacrificing accuracy on small scales, while retaining accuracy on large scales by exactly solving the linear perturbation theory equations.
The fiducial cosmology used to generate the simulations is listed as the first column in Table \ref{tab:powerspectrummodels}.

The mock catalogues were generated by taking a simulation snapshot at redshift $z=0.1$ (close to the effective redshift of 6dFGS) and populating it using a halo occupation distribution (HOD) model.
The HOD is informed by the number density function $n(z)$ and projected correlation function $w_p(r_p)$ of 6dFGS, and allocates both central and satellite galaxies to the N-body halos.
The random catalogues were populated by drawing Monte Carlo samples from the 6dFGS selection function, which accounts for both the angular and redshift distribution of galaxies in the survey.  
We apply our method to ten mock catalogues, which helps us assess the reliability of our method.

We take several additional steps to refine the ten mocks. 
Our logarithmic distance ratio sample is obtained by taking the 8885 most massive centrals in the 6dFGSv redshift range $z \leq 0.057$ and  converting the given peculiar velocity into a logarithmic distance ratio (via Eq. \ref{eq:vptoeta}).
We add mock measurement uncertainties to the logarithmic distance ratios by first drawing a random offset $\eta_{\rm offset}$ from the normal distribution $\mathcal{N}(0, \sigma_{\rm{obs}}=0.1)$, where $\sigma_{\rm obs}$ represents the typical level of uncertainty in the 6dFGSv logarithmic distance ratios.
We then modify the true logarithmic distance ratio through $\eta_{\rm modified} = \eta_{\rm true} + \eta_{\rm offset}$ and set the observed uncertainty to $\sigma_{\rm{obs}}$. 
Our galaxy overdensity sample is selected from both central and satellite galaxies below redshift $z=0.1$ (matching the redshift limit of our 6dFGS galaxy redshift sample).
When calculating the galaxy overdensities, we use a slightly different method for calculating $N_{\rm{exp}}$ compared to the 6dFGS data, which we estimate using the average $n(z)$ function from the random catalogues associated with the mock catalogues.
We did not use this method for the 6dFGS data because its angular selection function and number density function are not explicitly separable, hence the use of random catalogues determined from the luminosity function (see above).
Finally, the data mock catalogues are gridded at the same length-scales as the 6dFGS data: $L_{\delta_g} = 30$\mpch \ and $L_{\eta} = 20$\mpch.

\section{Theory and Methodology}\label{sec:theory_methodology}
As this work builds on that presented by \cite{Adams2017}, we note that the interested reader may refer back to that paper for the foundational information, as well as more depth, for the following sections.

\subsection{Likelihood Model} \label{subsec:likelihood_model}
In this study, we aim to extract constraints on the growth rate of structure by modelling the auto- and cross-covariance matrices of the peculiar velocity and galaxy overdensity fields. 
This is primarily done by constructing and evaluating a likelihood function, which describes the probability of observing the data given our model. 
This then informs the posterior probability distribution when we include known information in the form of the prior through Bayes' theorem.
The likelihood is a function of the data $\bm{\Delta}$ and the model parameters $\bm{\phi}$:
\begin{align}
\mathcal{L} &=  \frac{1}{\sqrt{(2\pi)^N |\mathbfss{C}(\bm{\phi})|}}\exp\left(-\frac{1}{2} \bm{\Delta}^T\mathbfss{C}(\bm{\phi})^{-1}\bm{\Delta}\right), \label{eq:likelihoodeq}
\end{align}
where $\mathbfss{C}(\bm{\phi})$ is the model covariance matrix and $N$ is the length of the data vector. 
In this method, the data vector contains $N_{\delta}$ galaxy overdensities $\bm{\delta}_g = (\delta_{g_1}, \delta_{g_2}, ..., \delta_{g_{N_{\delta}}})$, and $N_v$ peculiar velocities $\bm{v}_p = (v_{p_1}, v_{p_2}, ..., v_{p_{N_v}})$, such that it has length $N = N_{\delta} + N_{v}$. 
The model covariance matrix may be expressed using four submatrices, which are the auto- and cross-covariance matrices for the galaxy overdensity and peculiar velocity measurements:
\begin{align}
\mathbfss{C} = \begin{pmatrix}
\mathbfss{C}_{\delta \delta}  \ \mathbfss{C}_{\delta v} \\
\mathbfss{C}_{v \delta} \ \mathbfss{C}_{v v} \end{pmatrix}.
\end{align}

\subsection{Covariance Model} \label{subsec:cov_model}
The entries of the covariance matrix are determined by modelling the correlation between any two entries of the data vector. 
We begin with the expressions for the galaxy overdensity and peculiar velocity fields in Fourier space.

The observed galaxy overdensity field in redshift space is modelled as
\begin{align}
\tilde{\delta}_g^s(\bm{k}) &= [b\tilde{\delta}_m(\bm{k}) + f\mu^2\tilde{\theta}(\bm{k})]D_g(k,\mu, \sigma_g), \label{eq:rsd_densitydamp}
\end{align}
where $b$ is the galaxy bias in real space, $\delta_m$ is the matter overdensity field, $f$ is the growth rate of structure, $\mu \equiv \hat{\bm{k}}\cdot\hat{\bm{d}}$ is the angle between the wavevector $\bm{k}$ and the line-of-sight $\bm{d}$ and $\theta$ is the velocity divergence field.
The additional term 
\begin{align}
D_g(k,\mu, \sigma_g) &= e^{-(k\mu\sigma_g)^2/2} \label{eq:rsd_densitydamp_function}
\end{align}
is the damping due to the ``fingers-of-God'' effect, modelled by \cite{Peacock1994}.
Here, $\sigma_g$ is in units of \mpch; it characterises the strength of the damping and is related to the pairwise velocity dispersion. 

The observed peculiar velocity field is modelled as
\begin{align}
\tilde{v}_p(\bm{k}) &= -iaHf\frac{\mu}{k}\tilde{\theta}(\bm{k}) D_u(k, \sigma_u),  \label{eq:vkmudamp}
\end{align}
where $i$ is the imaginary unit, $a$ is the dimensionless scale factor and $H$ is the Hubble constant in units of \hubbleunit.
The additional term
\begin{align}
D_u(k, \sigma_u) &=  \frac{\sin(k\sigma_u)}{k\sigma_u} \label{eq:vkmudamp_function}
\end{align}
is the damping function introduced by \cite{Koda2014}, where $\sigma_u$ characterises the strength of the damping and is in units of \mpch.

One may obtain expressions for the correlation functions (and hence the complete model covariance), by Fourier transforming the corresponding power spectra.
The anisotropic auto- and cross-power spectra for the galaxy overdensity and peculiar velocity fields are
\begin{align}
P_{gg}(k,\mu) &= \begin{aligned}[t]b^2[P_{mm}(k) &+ 2r_g\beta\mu^2P_{m\theta}(k) \\&+ \beta^2\mu^4P_{\theta\theta}(k)]D_g^2(k,\mu,\sigma_g), \end{aligned} \label{eq:gganisotropicps} \\
P_{gv}(k,\mu) &=\begin{aligned}[t]\frac{iaHfb\mu}{k} [r_g P_{m\theta}(k) &+\beta\mu^2 P_{\theta\theta}(k)]\\& D_g(k,\mu,\sigma_g)D_u(k,\sigma_u), \end{aligned} \label{eq:gvanisotropicps} \\
P_{vg}(k,\mu) &=\begin{aligned}[t]\frac{-iaHfb\mu}{k} [r_g P_{m\theta}(k) &+\beta\mu^2 P_{\theta\theta}(k)]\\& D_g(k,\mu,\sigma_g)D_u(k,\sigma_u), \end{aligned}  \label{eq:vganisotropicps} \\
P_{vv}(k,\mu) &= \left(\frac{aHf\mu}{k}\right)^2 P_{\theta\theta}(k) D_u^2(k,\sigma_u). \label{eq:vvanisotropicps}
\end{align}
where we have introduced the cross-correlation coefficient $r_g$. 
This parameter allows for a more detailed galaxy bias relation, and is commonly used in RSD modelling \citep[e.g.][]{Dekel1999,Burkey2004,Blake2011b,Koda2014}. 
It is defined such that it modifies the galaxy bias only for the galaxy-matter cross-power spectrum.
Given that $\delta_m = \theta$ on linear scales, we have applied the cross-correlation coefficient to the cross-power spectrum $P_{g\theta} = bP_{m\theta}$ in our RSD equations.

Our method is structured such that we may vary two of $f$, $b$ and $\beta$ as free parameters.
However, recomputing the entire covariance matrix model for a new set of parameters is computationally expensive.
Instead, we break up each covariance such that the subsequent component matrices can be directly scaled by the free parameters.
The scaled components are then summed to get the complete model covariance.
We present a summary of the covariance equations below and provide the full derivation and equations for the components in Appendix \ref{sec:appendix}.

For the galaxy overdensity auto-covariance:
\begin{align}
\mathbfss{C}_{\delta \delta} = \frac{b^2}{2\pi^2}(\mathbfss{C}_{\delta \delta, \beta^0} + 2r_g\beta\mathbfss{C}_{\delta\delta, \beta^1} + \beta^2\mathbfss{C}_{\delta \delta, \beta^2}). \label{eq:main_ggbetasum}
\end{align}
Each of these components may then be expressed as the Fourier transform of the corresponding power spectrum.
The Fourier transform is over angle $\mu$ and wavenumber $k$, so we evaluate the angular component analytically using multipole expansion, where the multipole orders are given by $\ell$ (see Appendix \ref{sec:appendix}).
This leads to us expressing the components as an integral over $k$:
\begin{align}
\mathbfss{C}_{\delta \delta, \beta^0} &= \int k^2 P_{mm}(k) \sum_{\ell \in 0,2,4}\mathbfss{K}_{\delta \delta, \beta^0, \ell} \ dk, \\
\mathbfss{C}_{\delta \delta, \beta^1} &= \int k^2 P_{m\theta}(k) \sum_{\ell \in 0,2,4}\mathbfss{K}_{\delta \delta, \beta^1, \ell} \ dk, \\
\mathbfss{C}_{\delta \delta, \beta^2} &= \int k^2 P_{\theta\theta}(k) \sum_{\ell \in 0,2,4}\mathbfss{K}_{\delta \delta, \beta^2, \ell} \ dk,  
\end{align}
where $\mathbfss{K}_{\delta \delta, \beta, \ell}$ are matrices containing the integrands for each order of $\beta$ and $\ell$, with their functional forms expressed in Eq. \ref{eq:K_dd_begin} to \ref{eq:K_dd_end}.

For the peculiar velocity auto-covariance:
\begin{align}
\mathbfss{C}_{v v} = \frac{(aHf)^2}{2\pi^2}(\mathbfss{C}_{vv, \beta^0}),
\end{align}
where
\begin{align}
\mathbfss{C}_{vv, \beta^0} = \int P_{\theta\theta}(k) D^2_u(k,\sigma_u) \sum_{\ell \in 0,2}\mathbfss{K}_{vv, \ell} \ dk,
\end{align}
where $\mathbfss{K}_{vv, \ell}$ are matrices containing the integrands for each order of $\ell$, with their functional forms expressed in Eq. \ref{eq:K_vv_begin} to \ref{eq:K_vv_end}.

Finally, for the cross-covariance:
\begin{align}
\mathbfss{C}_{v \delta} = \frac{aHfb}{2\pi^2}(r_g\mathbfss{C}_{v \delta, \beta^0} + \beta\mathbfss{C}_{v \delta, \beta^1}), \label{eq:dvcov_components}
\end{align}
where
\begin{align}
\mathbfss{C}_{v \delta, \beta^0} &= \int k P_{\theta m}(k) D_u(k,\sigma_u) \sum_{\ell \in 1,3} \mathbfss{K}_{v \delta, \beta^0, \ell} \ dk, \\
\mathbfss{C}_{v \delta, \beta^1} &= \int k P_{\theta \theta}(k) D_u(k,\sigma_u) \sum_{\ell \in 1,3} \mathbfss{K}_{v \delta, \beta^1, \ell} \ dk,
\end{align}
where $\mathbfss{K}_{\delta \delta, \beta, \ell}$ are matrices containing the integrands for each order of $\beta$ and $\ell$, with their functional forms expressed in Eq. \ref{eq:K_dv_begin} to \ref{eq:K_dv_end}.

\subsection{Model Modifications}
In \cite{Adams2017}, we introduced several modifications to the covariance model to better capture the data.
We refer the reader to sections 3.3, 3.4 and 3.7 in that work, and provide a brief overview of the key modifications below.

\subsubsection{Modelling the Logarithmic Distance Ratio}
\cite{Springob2014} showed that peculiar velocities measured from 6dFGS have log-normal, rather than Gaussian, uncertainty distributions, making them unsuitable for our likelihood model.
However, the use of the Fundamental Plane method in that work also allows the peculiar velocity measurements to be written as a logarithmic distance ratio, defined as $\eta \equiv \log_{10}[D(z_{\text{obs}})/D({z_{H}})]$, which does have a Gaussian distribution.
Here, $D(z_{\text{obs}})$ is the comoving distance inferred from the observed redshift (including the peculiar velocity component) and $D(z_{H})$ is the true comoving distance, estimated using the Fundamental Plane method \citep[for more detail, see][]{Springob2014}.

We use a conversion factor to write the model for peculiar velocity in terms of the logarithmic distance ratio:
\begin{align}
\xi(z_{\text{obs}}) &= \frac{1}{\ln(10)} \frac{1 + z_{\text{obs}}}{D(z_{\text{obs}})H(z_{\text{obs}})}. \label{eq:veltoetaconversiondef}
\end{align}
This relation has been previously used for 6dFGS by both \cite{Johnson2014} and \cite{Adams2017}, following work by \cite{Hui2006} on how peculiar velocities affected supernova magnitudes.
For notational simplicity, we do not explicitly state the redshift dependence but note that it is implied by the peculiar velocity that is modified through
\begin{align}
\eta = \xi v_p. \label{eq:vptoeta}
\end{align}

We note that each element of the model covariance corresponds to observations at two points in space, arbitrarily labelled $\bm{x}_s$ and $\bm{x}_t$ (see fig. 2 of \citealt{Adams2017}).
Consequently, when discussing how individual covariance elements are modified, we use the notation $C(\bm{x}_s, \bm{x}_t)$.
In the case of the conversion factor $\xi$, the covariance equations are
\begin{align}
C_{\eta \eta} (\bm{x}_s, \bm{x}_t) &= \xi^2 C_{v v} (\bm{x}_s, \bm{x}_t), \\
C_{\delta \eta} (\bm{x}_s, \bm{x}_t) &= \xi C_{\delta v} (\bm{x}_s, \bm{x}_t),\\
C_{\eta \delta} (\bm{x}_s, \bm{x}_t) &= \xi C_{v \delta} (\bm{x}_s, \bm{x}_t).
\end{align}

\subsubsection{Accounting for Gridding} \label{subsubsection:gridding}
In this method, gridding allows us to calculate the galaxy overdensity, smooth over non-linear effects, and reduce the computation time required to evaluate the likelihood equation, which scales with the size of the data vector and covariance.
We follow the modelling approach outlined by \cite{Abate2008}.

Importantly, gridding reduces small-scale power, which we account for by multiplying our model power spectra by a gridding window function $\Gamma$, where
\begin{align}
\Gamma(k,L) &= \left\langle \frac{8}{L^3} \frac{\sin  \left (k_x \tfrac{L}{2}\right )}{k_x} \frac{\sin  \left (k_y \tfrac{L}{2}\right )}{k_y} \frac{\sin  \left (k_z \tfrac{L}{2}\right )}{k_z} \right\rangle_{\bm{k} \in k}.
\end{align} 
Here, $L$ is the length of the grid cell in \mpch, and the average is applied to all $\bm{k}$ vectors that have magnitude $k$. Since we may use different gridding sizes for peculiar velocities and overdensities, we define $\Gamma_{\delta}(k) = \Gamma(k,L_{\delta})$ and $\Gamma_{v}(k) = \Gamma(k,L_{v})$.

The smoothing functions are then included in each covariance integrand, such that the covariance components are modified.
For example,
\begin{align}
\mathbfss{C}'_{\delta \delta, \beta^0} &= \int k^2 P_{mm}(k) \sum_{\ell \in 0,2,4}\mathbfss{K}_{\delta \delta, \beta^0, \ell}\ \Gamma_{\delta}^2(k) \ dk, \\
\mathbfss{C}'_{v v, \beta^0} &= \int P_{\theta\theta}(k) D^2_u(k,\sigma_u) \sum_{\ell \in 0,2}\mathbfss{K}_{vv, \ell}\ \Gamma_{v}^2(k) \ dk,\\
\mathbfss{C}'_{v \delta, \beta^0} &= \int k P_{\theta m}(k) D_u(k,\sigma_u) \sum_{\ell \in 1,3} \mathbfss{K}_{v \delta, \beta^0, \ell}\ \Gamma_{\delta}(k)\Gamma_{v}(k) \ dk,
\end{align}
and so on for each order of $\beta$.
The modified covariances may then be summed to give the total smoothed covariances $\mathbfss{C}'_{\delta \delta}$, $\mathbfss{C}'_{vv}$ and $\mathbfss{C}'_{v\delta}$.

In addition to the smoothing, \cite{Abate2008} also correct the velocity auto-covariance after taking the average value in each cell by introducing a shot noise component. 
For $N_s$ values of $v$ in a cell at position $\bm{x}_s$, the shot noise contribution can be calculated as
\begin{align}
\sigma_{\rm sn}^2(\bm{x}_s, \bm{x}_t) &= \frac{C_{v v}(\bm{x}_s,\bm{x}_t) - C'_{v v}(\bm{x}_s,\bm{x}_t)}{N_s} \delta_{st}, \label{eq:sigmacorr}
\end{align}
where $\delta_{st}$ is the Kronecker delta, such that $\sigma_{\rm sn}^2$ only appears on the diagonal. This term behaves similarly to an error term, so we incorporate it into the model in the next section.

\subsubsection{Inclusion of Error Terms}
We incorporate a number of error terms; assuming that the error in any given data point is independent, errors only appear along the diagonal of the covariance matrix. 

For the measured value of $\eta$ at a given position $\bm{x}_i$, we include the uncertainty in the measurement from the Fundamental Plane $\sigma_{\text{obs}}(\bm{x}_i)$ \citep[as measured by ][]{Springob2014}, the shot noise from averaging the value in each cell $\sigma_{\rm sn}$ (see Eq. \ref{eq:sigmacorr}), and a stochastic velocity term to account for the breakdown of linear theory $\sigma_v$.
The contribution to the logarithmic distance ratio auto-covariance will then be
\begin{align}
&\sigma^2_{\eta \eta}(\bm{x}_s, \bm{x}_t) = \sigma_{\rm obs} (\bm{x}_s)\sigma_{\rm obs}(\bm{x}_t)\delta_{st} + \xi^2\sigma_{\rm sn}^2(\bm{x}_s, \bm{x}_t) + \xi^2\sigma^2_{v}\delta_{st},
\end{align}
where $\delta_{st}$ is the Kronecker delta, ensuring that error terms only affect the diagonal of the covariance matrix. 
The logarithmic distance ratio auto-covariance becomes
\begin{align}
&C_{\eta \eta}^{\rm err} (\bm{x}_s, \bm{x}_t) = C'_{\eta \eta}(\bm{x}_s, \bm{x}_t) + \sigma^2_{\eta \eta}(\bm{x}_s, \bm{x}_t).
\end{align}

For the measured value of $\delta_g$ at a given position $\bm{x}_i$, we include the shot noise contribution $\sigma_{\delta_g}$ (discussed in Section \ref{subsec:data}) such that the error contribution to the galaxy overdensity auto-covariance will then be
\begin{align}
&\sigma^2_{\delta_g \delta_g}(\bm{x}_s, \bm{x}_t) = \sigma_{\delta_g} (\bm{x}_s)\sigma_{\delta_g} (\bm{x}_t)\delta_{st}.
\end{align}
The galaxy overdensity auto-covariance becomes
\begin{align}
&C_{\delta_g \delta_g}^{\rm err}  (\bm{x}_s, \bm{x}_t) = C'_{\delta_g \delta_g} (\bm{x}_s, \bm{x}_t) + \sigma^2_{\delta_g \delta_g}(\bm{x}_s, \bm{x}_t).
\end{align}

\subsubsection{Accounting for Redshift-Dependent Galaxy Bias} \label{subsubsec:alphab}
As discussed in Section \ref{subsec:data}, the bias of the galaxy overdensity sample will increase with redshift for our magnitude-limited sample.
It is important to account for this effect because the covariance model is a function of the effective bias over the redshift range, rather than the bias as a function of redshift.
The amplitude of the cross-correlation is proportional to the linear bias factor, but the value of the bias will be determined by the overdensities that directly influence the peculiar velocities, rather than the entire galaxy overdensity sample.
Consequently, we naturally expect that the cross-correlation of peculiar velocities with a lower-bias overdensity sample will not be as strong as that with a higher-bias sample.
Given that the peculiar velocity sample is limited to a lower redshift, we anticipate that the effective bias probed by the cross-correlation will be lower than that probed by the galaxy overdensity auto-correlation.

We propose a simple modification that allows for the cross-covariance to have a lower effective bias value than the galaxy overdensity auto-covariance.
Until now, the cross-covariance has used the same galaxy bias value as the galaxy overdensity auto-covariance, $b$. 
We modify this using a scaling parameter $\alpha_b$, such that in the cross-covariance
\begin{align}
b &\rightarrow \alpha_b b,
\end{align}
giving the overall transformation
\begin{align}
C'_{\delta \eta}(\bm{x}_s, \bm{x}_t) &\rightarrow \alpha_b C'_{\delta \eta}(\bm{x}_s, \bm{x}_t) \label{eq:modified_dv_alphab},\\
C'_{\eta \delta}(\bm{x}_s, \bm{x}_t) &\rightarrow \alpha_b C'_{\eta \delta}(\bm{x}_s, \bm{x}_t), \label{eq:modified_vd_alphab}
\end{align}
while $\mathbfss{C}'_{\delta \delta}$ remains unchanged.

\subsection{Evaluating the Likelihood}
We now cover the key steps required to evaluate our model covariance and the likelihood function.

\subsubsection{Generating the Fiducial Power Spectra}
To evaluate the covariance model, we must provide model power spectra for our fiducial cosmology.
We note that $f$ and $b$ are degenerate with the amplitude of the fiducial power spectra, parametrized by $\sigma_{8}$. 
Consequently, our analysis constrains $f\sigma_{8}$ and $b\sigma_{8}$ and we divide the power spectra by their fiducial $\sigma_8$ value to normalise the amplitude.

We generate the matter power spectrum $P_{mm}(k)$ from the Code for Anisotropies in the Microwave Background \citep[\texttt{CAMB};][]{Lewis2000, Lewis2011}, utilising the non-linear corrections from Halofit. 
For the velocity divergence power spectrum $P_{\theta\theta}(k)$ and the cross power spectrum $P_{m\theta}(k) = P_{\theta m}(k)$ we use \texttt{velMPTbreeze}, an extension to \texttt{MPTbreeze} \citep{Crocce2012} for calculating velocity power spectra. 

We use a number of cosmological parameter sets in our analysis, which are listed in Table \ref{tab:powerspectrummodels}.
When working with the COLA mocks, we use the same cosmological parameters that were used to generate the simulations, which allows us to test whether our method recovers the expected value for $f\sigma_8$. 
For the 6dFGS data, we use the Planck 2015 cosmological parameter values \citep{PlanckCollaboration2015a} for our key results.
To test how the choice of cosmological parameters affects our results, we also use cosmological parameters from the Wilkinson Microwave Anisotropy Probe (WMAP) five-year data \citep{Komatsu2009}, and those from the Planck 2018 data \citep{PlanckCollaboration2018}.  

\begin{table}
	\centering
	\caption[]{Cosmological parameters for the four cosmologies used in this analysis. The top section shows the 6 base parameters for standard $\Lambda$CDM: physical baryon density; physical dark matter density; reduced Hubble constant; scalar spectral index; scalar amplitude (with pivot point $k_0=0.002$\hmpc \ for the WMAP cosmology and $k_0=0.05$\hmpc \ for the mock and Planck cosmologies); and reionization optical depth. The bottom section shows the fiducial $\sigma_8$ for each cosmology, which is a derived parameter.}
	\label{tab:powerspectrummodels}
	\begin{tabular}{lllll} % four columns, alignment for each
		\hline
		& COLA Mocks & Planck 2015 & WMAP & Planck 2018 \\
		\hline
		$\Omega_bh^2$ & 0.02210 & 0.02222 & 0.02273 & 0.02212\\ 
		$\Omega_ch^2$ & 0.1166 & 0.1197 & 0.1099 & 0.1206\\
		$h$ & 0.68 & 0.6731 & 0.719 & 0.6688\\
		$n_s$ & 0.96 & 0.9655& 0.963 & 0.9626\\
		$A_s$ & 2.215$\times10^{-9}$ & 2.195$\times10^{-9}$ &  2.41$\times10^{-9}$ & 2.092$\times10^{-9}$\\
		$\tau$ & 0.09 & 0.078 & 0.087& 0.0522\\
		\hline
		$\sigma_8^{\rm fid}$ & 0.82 & 0.8417 & 0.7931 & 0.8118\\
		\hline
	\end{tabular}
\end{table}

\subsubsection{Integration Bounds} \label{subsubsec:integration_bounds}
As part of evaluating the covariance equations, we must specify the bounds for the integral over $k$.
We use the same bounds of $k_{\text{min}} = 0.0025$\hmpc \ and $k_{\text{max}} = 0.15$\hmpc \ as those from \cite{Adams2017}, but note that we vary $k_{\text{max}}$ during the analysis to understand how it affects our results.

\cite{Adams2017} found evidence that there was a significant contribution to the galaxy overdensity auto-covariance beyond $k_{\text{max}} = 0.15$\hmpc, so introduced an additional integral, that ranged from $k_{\text{max}}$ to $k_{\text{add}} = 1.0$\hmpc.
We choose to keep this component in the model, and test whether its inclusion is justified when working with the COLA mocks.
In doing so, we introduce the bias parameter that scales the additional integral as $b_{\text{add}}\sigma_{8}$, noting that it behaves similarly to the linear galaxy bias $b\sigma_{8}$.

\subsubsection{MCMC Sampling} \label{subsubsec:mcmc_sampling}
In this analysis, we constrain our free parameters through a Markov chain Monte Carlo (MCMC) method.
We use \texttt{emcee} \citep{Foreman-Mackey2013}, which is a \texttt{Python} implementation of the affine-invariant ensemble sampler for MCMC proposed by \cite{Goodman2010}.
The MCMC chains in our analysis were run with 500 walkers taking 800 steps, which equates to 400,000 samples of our parameter space.
We discard the first 150 steps as burn-in and confirm that the chains have converged using the Gelman-Rubin statistic $\hat{R}$ \citep{Gelman1992}.
If the Gelman-Rubin statistic is close to one, then the chains have converged to the posterior distribution; we use the condition that $\hat{R} - 1$ must be less than 0.05 to satisfy convergence.
This convergence test is already implemented as part of the \texttt{ChainConsumer} analysis package \citep{Hinton2016}, which we use to analyse all of our \texttt{emcee} chains.

The likelihood function (Eq. \ref{eq:likelihoodeq}) is evaluated at each step for each walker.
The covariance is effectively inverted by applying the Linear Algebra PACKage (LAPACK) Cholesky solver to the equation $\mathbfss{C}\bm{\Upsilon} = \bm{\Delta}$, which yields $\bm{\Upsilon} = \mathbfss{C}^{-1}\bm{\Delta}$.
We use the \texttt{Python} implementation of LAPACK available through the \texttt{SciPy Linear Algebra} package.
The exponent of the likelihood equation is obtained by multiplying $\bm{\Upsilon}$ by $-\tfrac{1}{2}\bm{\Delta}^T$. 

We note that the damping functions for RSD (see Eq. \ref{eq:rsd_densitydamp_function} and \ref{eq:vkmudamp_function}) introduce two parameters ($\sigma_g$ and $\sigma_u$) that cannot be varied as free parameters in the \texttt{emcee} runs. 
This is because they exist inside the integral over the wavenumber $k$ and varying them would involve a recalculation of the entire covariance matrix (which is currently computationally intractable) rather than the simple rescaling that comes from breaking the model covariance into components (see Section \ref{subsec:cov_model}).
Consequently, we fix these as part of the analysis but examine how different values affect the results.

\section{Simulation Results} \label{sec:simresults}

We wish to validate our method by testing whether our pipeline recovers the expected cosmology used to generate the 6dFGS mock catalogues (see Table \ref{tab:powerspectrummodels}). 
For the $\Omega_m$ and $\sigma_8$ values used to generate the mock catalogues, the expected growth rate of structure is $f\sigma_8 = 0.423$.

The nature of the covariance evaluation means that $k_{\rm{max}}$, $\sigma_u$ and $\sigma_g$ cannot be varied as free parameters in the analysis. 
Therefore, it is important to examine whether the choices we make for the values of these parameters affect the constraint on the growth rate. 
We start by evaluating the likelihood using only the galaxy overdensity and logarithmic distance ratio auto-covariances independently. 
Once optimal parameter values are established, we fix these and move on to evaluating the likelihood with the complete covariance, testing different values of $\alpha_b$.

\subsection{Galaxy Overdensity Auto-Covariance} \label{subsec:sims_ddcov}
We begin by establishing our best estimates for the fixed parameters used in the galaxy overdensity auto-covariance model. 
The $k$-range we fit over is controlled by $k_{\rm{max}}$, which we take to be $k_{\rm{max}} = 0.15$\hmpc \ (see Section \ref{subsubsec:integration_bounds}).
We also include the additional integral, parametrized by the nuisance parameter $b_{\rm{add}}\sigma_8$. 
We do not model RSD in the galaxy overdensity auto-covariance model beyond $k_{\rm{max}}$, making the non-linear covariance independent of the growth rate.
We set the damping due to pairwise velocities as $\sigma_g = 3.0$\mpch; this corresponds to a pairwise velocity dispersion of $300$\kms, which is a standard fiducial assumption \citep[e.g.][]{Peacock1994, Blake2018}. 
We set the strength of the peculiar velocity power spectrum damping as $\sigma_u = 13.0$\mpch, which is the preferred value from \cite{Koda2014}.
We take the galaxy cross-correlation coefficient to be $r_g = 1$, which corresponds to the linear bias model. 
We refer to this collection of model parameter values as our fiducial model for the galaxy overdensity auto-covariance.

\begin{figure}
	\includegraphics[width=\columnwidth]{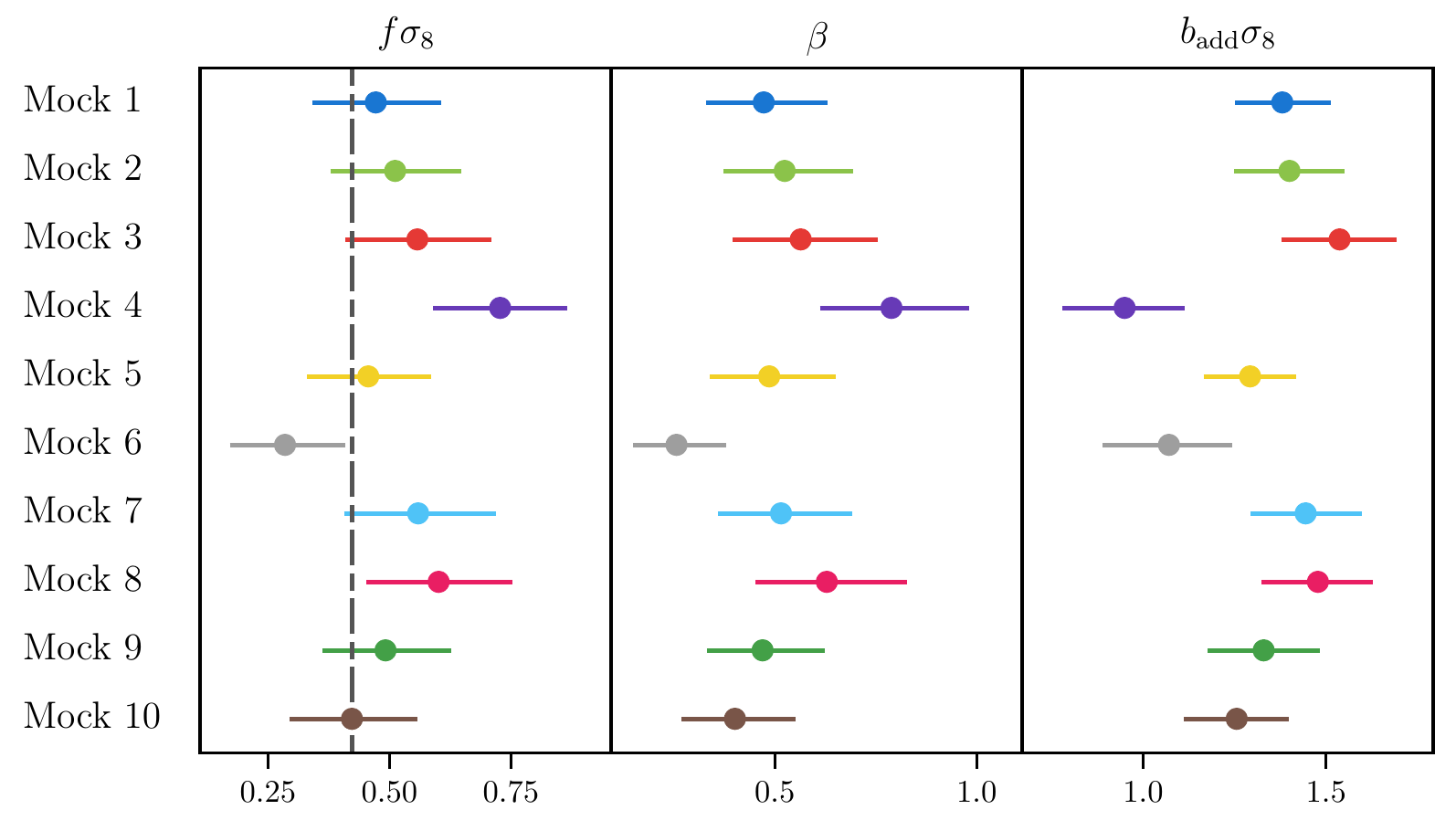}
	\caption{The median values and 68\% credible intervals of $f\sigma_8$, $\beta$ and $b_{\rm{add}}\sigma_8$ for ten 6dFGS mocks when using the galaxy overdensity auto-covariance. The expected value for $f\sigma_8$ is shown by the dashed vertical line.}
	\label{fig:sims_dd_allmocks_fiducial}
\end{figure}

We start by evaluating the likelihood for the galaxy overdensity measurements from ten 6dFGS mocks using our fiducial model. 
The marginalised constraints on the three free parameters for this model ($f\sigma_8$, $\beta$, $b_{\rm{add}}\sigma_8$) are given in Fig. \ref{fig:sims_dd_allmocks_fiducial}.
We note that the credible intervals for each parameter have roughly consistent sizes across all ten mocks, which shows that the precision of our method is robust. 
The fiducial $f\sigma_{8}$ value is recovered at the 1$\sigma$ level for seven of the ten mocks.
The relative positions of the $\beta$ credible intervals between each mock are similar to those for $f\sigma_8$, which indicates that the galaxy bias is consistent across the mocks.
Given that the mocks are independent, we also calculate the mean and error in the mean for the growth rate, finding $f\sigma_{8, \text{mean}} = 0.51 \pm 0.04$.
We note that the mean growth rate is not consistent with the fiducial $f\sigma_8$ value at the 1$\sigma$ level, likely due to the bias towards higher values visible in Fig. \ref{fig:sims_dd_allmocks_fiducial}.
This could be due to the choice of fiducial value for $\sigma_g$, which we discuss later in this section.

\begin{figure}
	\includegraphics[width=\columnwidth]{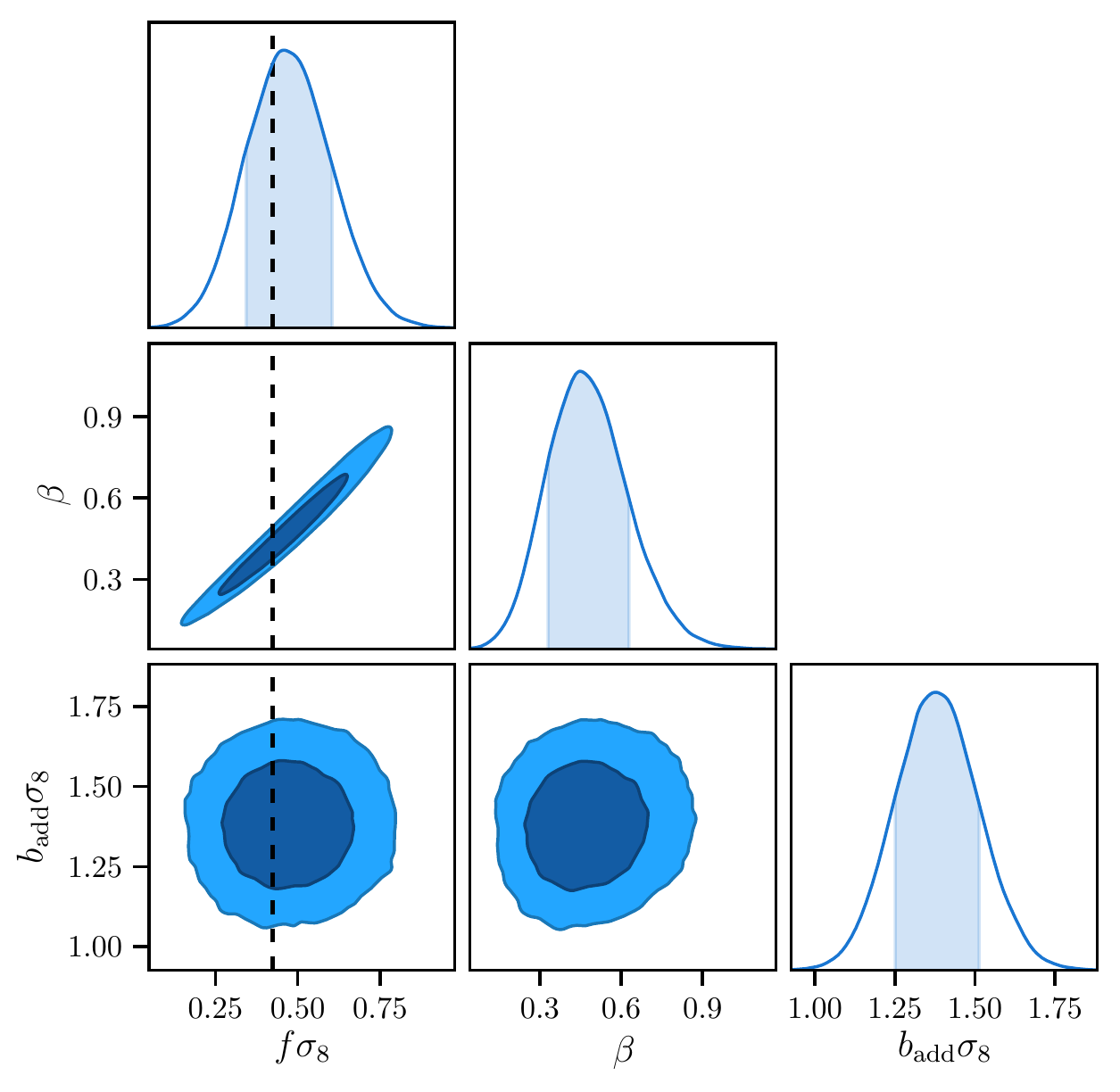}
	\caption{The posteriors of our free parameters for Mock 1 when using the galaxy overdensity auto-covariance. The shaded area of each marginalised posterior represents the 68\% credible interval. The dark shading on each contour indicates the 68\% credible region and the lighter shading indicates the 95\% credible region. The expected value for $f\sigma_8$ is shown by the dashed vertical line.}
	\label{fig:sims_dd_mock1_fiducial}
\end{figure}

To get an appreciation for the degeneracies between the three parameters, we show the corner plot for Mock 1, which we take as a representative sample, in Fig. \ref{fig:sims_dd_mock1_fiducial}.
From this, we can see that we recover the expected growth rate of structure at the 1$\sigma$ level. 
The contours indicate that there is a slight degeneracy between $\beta$ and $b_{\rm{add}}\sigma_8$, which is to be expected given that \cite{Adams2017} saw similar behaviour between $b_{\rm{fit}}\sigma_8$ and $b_{\rm{add}}\sigma_8$ (see fig. 5 in that work). 
We note that there is minimal degeneracy between $f\sigma_8$ and $b_{\rm{add}}\sigma_8$. 
Finally, the tight slope of the $\beta$-$f\sigma_8$ contour indicates that a single effective galaxy bias value is a reasonable model when fitting to linear scales.

Given that several model parameters are fixed, it's important to further investigate how varying these impacts the results, and consequently, whether our choice of parameters for the fiducial model is justified.
Before proceeding, we note that the maximum likelihood for our fiducial model corresponds to $\chi^2 = 1581.88$ ($\chi^2/\text{dof} = 0.97$), indicating a reasonable fit.

\begin{figure}
	\includegraphics[width=\columnwidth]{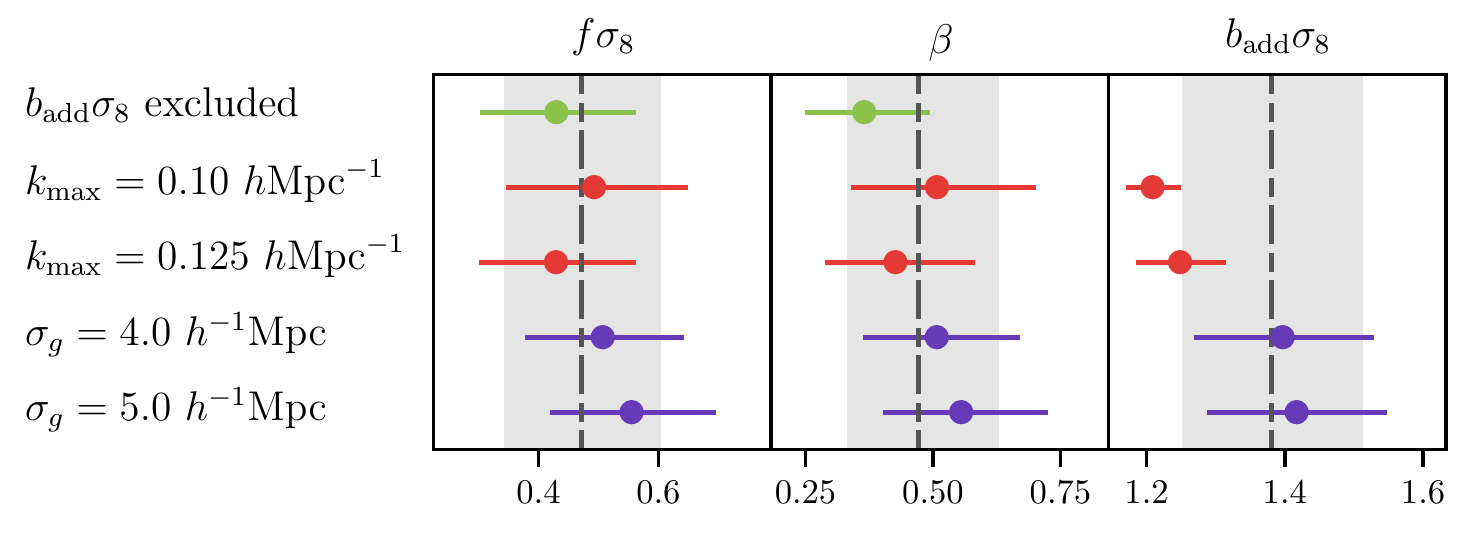}
	\caption{The median values and 68\% credible intervals of $f\sigma_8$, $\beta$ and $b_{\rm{add}\sigma_8}$ for various tests on Mock 1 when using the galaxy overdensity auto-covariance. The dashed lines and grey shaded regions represent the median value and 68\% credible interval for our fiducial model: $b_{\rm{add}}\sigma_8$ included, $k_{\rm{max}}=0.15$\hmpc, $\sigma_g=3.0$\mpch.}
	\label{fig:sims_dd_mock1_systematics}
\end{figure}

Firstly, we assess whether we are justified in using $b_{\rm{add}}\sigma_8$ as a nuisance parameter.
We run the covariance model while excluding $b_{\rm{add}}\sigma_8$ and present the median values and 68\% credible intervals for $f\sigma_8$ and $\beta$ relative to those from the fiducial model in Fig. \ref{fig:sims_dd_mock1_systematics}.
The maximum likelihood when excluding $b_{\rm{add}}\sigma_8$ from the model corresponds to $\chi^2 = 1680.45$ ($\chi^2/\text{dof} = 1.03$).
We can see that excluding $b_{\rm{add}}\sigma_8$ lowers both $f\sigma_8$ and $\beta$. 
The effect on $\beta$ is greater because the additional power that was being contributed from the non-zero $b_{\rm{add}}\sigma_8$ has been transferred to the linear-scale bias, lowering $\beta$ for a fixed growth rate.
We determine that the difference in $\chi^2$ between including and excluding $b_{\rm{add}}\sigma_8$, interpreted using the reduced Akaike information criterion \citep[AICc; see][]{Burnham2004}, is substantial evidence for keeping $b_{\rm{add}}\sigma_8$ as a model parameter.

Next, we test how the value of $k_{\rm{max}}$ affects the constraints.
We run the covariance model while setting $k_{\rm{max}} =0.10$, $0.125$\hmpc and present the median values and 68\% credible intervals for $f\sigma_8$, $\beta$ and $b_{\rm{add}}\sigma_8$ relative to those from the fiducial model in Fig. \ref{fig:sims_dd_mock1_systematics}.
The maximum likelihood when setting $k_{\rm{max}} = 0.10$\hmpc \ corresponds to $\chi^2 = 1578.90$ ($\chi^2/\text{dof} = 0.97$).
The maximum likelihood when setting $k_{\rm{max}} = 0.125$\hmpc \ corresponds to $\chi^2 = 1581.35$ ($\chi^2/\text{dof} = 0.97$).
Given the minimal impact of this choice on the measured growth rate and likelihood, we choose to keep our fiducial model value of $k_{\rm{max}} = 0.15$\hmpc as it maximises the range over which $f\sigma_8$ is fit.

Finally, we also test how the value of $\sigma_g$ affects the constraints.
We run the covariance model while setting $\sigma_g =4.0$, $5.0$\mpch \ and present the median values and 68\% credible intervals for $f\sigma_8$, $\beta$ and $b_{\rm{add}}\sigma_8$ relative to those from the fiducial model in Fig. \ref{fig:sims_dd_mock1_systematics}.
The maximum likelihood when setting $\sigma_g =4.0$\mpch \ corresponds to $\chi^2 = 1580.22$ ($\chi^2/\text{dof} = 0.97$).
The maximum likelihood when setting $\sigma_g =5.0$\mpch \ corresponds to $\chi^2 = 1579.86$ ($\chi^2/\text{dof} = 0.97$).
We can see that there is a systematic shift in both $f\sigma_8$ and $\beta$ in proportion to the value of $\sigma_g$, while $b_{\rm{add}}\sigma_8$ remains largely unaffected.
This is consistent with the fact that $\sigma_g$ controls the level of damping and that stronger damping will result in larger $f\sigma_8$ values as the covariance compensates, similar to the trade-off between $\beta$ and $b_{\rm{add}}\sigma_8$.
This effect has also been seen in other RSD studies \citep[see figure 4 in][]{Peacock2001}.
We also attempted to run tests using $\sigma_g = 2.0$, $1.0$\mpch.
Unfortunately, the numerical integration library we used to calculate the covariance matrix elements failed in both cases due to round-off errors.
Given the trend of $f\sigma_8$ with $\sigma_g$, we keep   $\sigma_g= 3.0$\mpch \ as our fiducial value.

Out of all three tests, Fig. \ref{fig:sims_dd_mock1_systematics} shows us that $\sigma_g$ has the largest effect on $f\sigma_8$ and that the value of $k_{\rm{max}}$ has the largest effect on $b_{\rm{add}}\sigma_8$.
For $f\sigma_8$, the various systematic tests all return median values that are within 1$\sigma$ of our fiducial case.
We later use these tests to quantify the systematic uncertainty for our parameter estimates when working with the 6dFGS data in Section \ref{subsec:systematicuncert}.

\subsection{Logarithmic Distance Ratio Auto-Covariance} \label{subsec:sims_vvcov}
Again, we begin by establishing our best estimates for the fixed parameters used in the logarithmic distance ratio auto-covariance model.
We again take $k_{\rm{max}}=0.15$\hmpc \ as the boundary to our fitted $k$-range. 
The damping identified by \cite{Koda2014} is implemented as a sinc function and parametrized by $\sigma_u$ (see Eq. \ref{eq:vkmudamp_function}). 
We take $\sigma_u=13.0$\mpch \ as our best estimate, given that \cite{Koda2014} found this to be the best fit to their simulations.
We refer to this collection of model parameter values as our fiducial model for the logarithmic distance ratio auto-covariance.

\begin{figure}
	\includegraphics[width=\columnwidth]{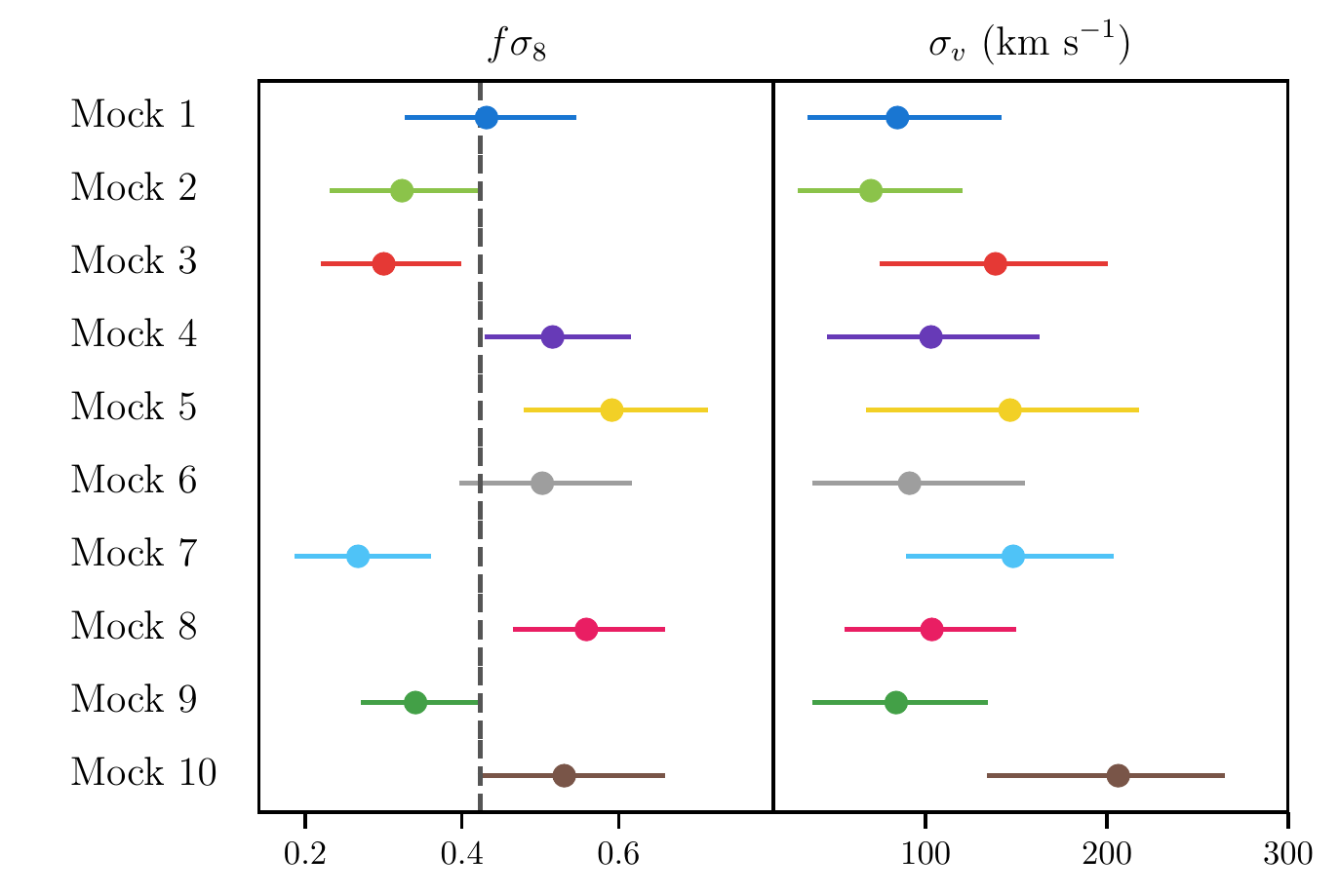}
	\caption{The median values and 68\% credible intervals of $f\sigma_8$ and $\sigma_v$ for ten 6dFGS mocks when using the logarithmic distance ratio auto-covariance. The expected value for $f\sigma_8$ is shown by the dashed vertical line.}
	\label{fig:sims_vv_allmocks_fiducial}
\end{figure}

As in the previous section, we examine the constraints for all ten mocks when using our best estimates. 
The marginalised constraints for the free parameters of this model ($f\sigma_8$, $\sigma_v$) are given in Fig. \ref{fig:sims_vv_allmocks_fiducial}.
As with the galaxy overdensity auto-covariance, the credible intervals for our free parameters are roughly consistent across all ten mocks.
The method does not appear to be biased; the ten mocks are evenly distributed around the expected recovery value for $f\sigma_8$, and the value is recovered at the 1$\sigma$ level in six of the ten mocks. 
Given that the mocks are independent, we also calculate the mean and error in the mean for the growth rate, finding $f\sigma_{8, \text{mean}} = 0.44 \pm 0.03$.
We note that the mean growth rate is consistent with the fiducial $f\sigma_8$ value at the 1$\sigma$ level.

\begin{figure}
	\includegraphics[width=\columnwidth]{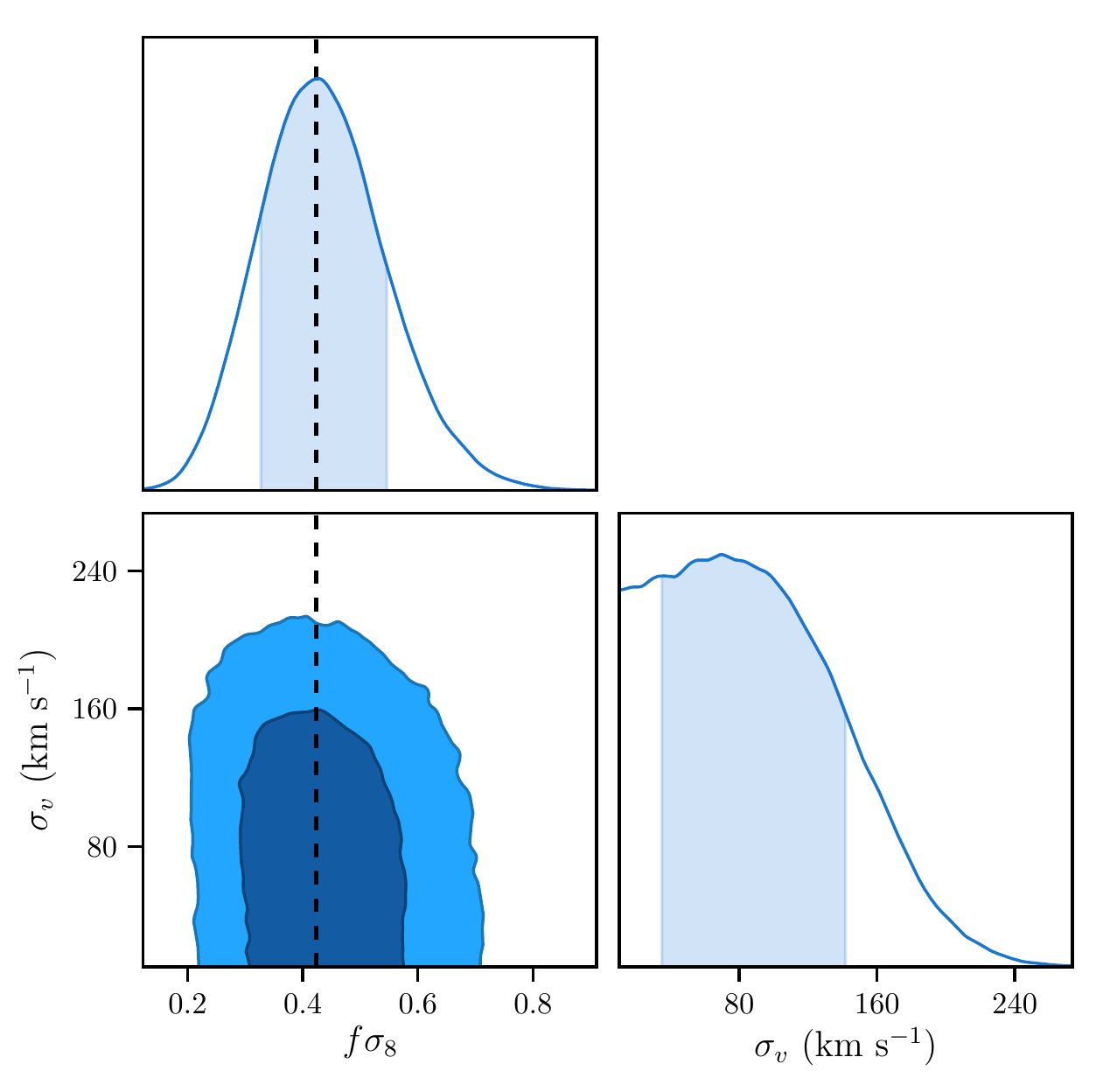}
	\caption{The posteriors of our free parameters for Mock 1 when using the logarithmic distance ratio auto-covariance. The shaded area of each marginalised posterior represents the 68\% credible interval. The dark shading on each contour indicates the 68\% credible region and the lighter shading indicates the 95\% credible region. The expected value for $f\sigma_8$ is shown by the dashed vertical line.}
	\label{fig:sims_vv_mock1_fiducial}
\end{figure}

We again take Mock 1 as a representative sample and present the corner plot for this mock in Fig. \ref{fig:sims_vv_mock1_fiducial}.
There appears to be minimal degeneracy between $f\sigma_8$ and $\sigma_v$, and the expected growth rate is well-recovered by the mock. 
The maximum likelihood for our fiducial model corresponds to $\chi^2 = 1054.78$ ($\chi^2/\text{dof} = 0.91$), indicating a reasonable fit.

\begin{figure}
	\includegraphics[width=\columnwidth]{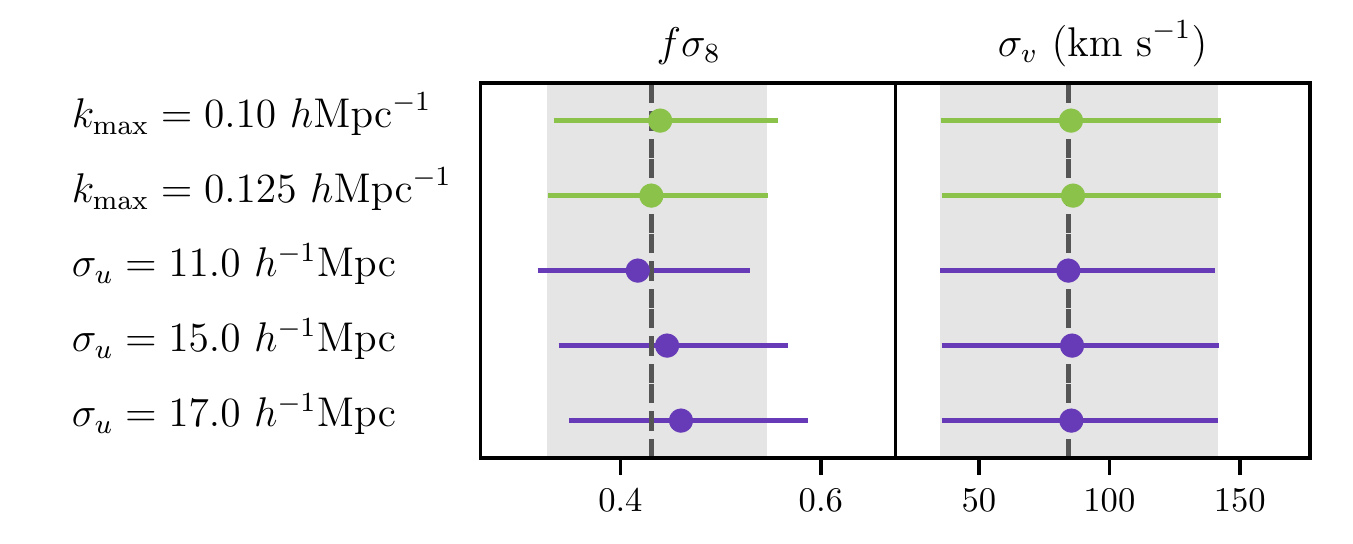}
	\caption{The median values and 68\% credible intervals of $f\sigma_8$ and $\sigma_v$ for various tests on Mock 1 when using the logarithmic distance ratio auto-covariance. The dashed lines and grey shaded regions represent the median value and 68\% credible interval for our fiducial model: $k_{\rm{max}}=0.15$\hmpc,  $\sigma_u=13.0$\mpch.}
	\label{fig:sims_vv_mock1_systematics}
\end{figure}

As in the previous section, we test how varying our fixed parameters alters the results, beginning with $k_{\rm{max}}$.
We run the covariance model while setting $k_{\rm{max}} =0.10$, $0.125$\hmpc \ and present the median values and 68\% credible intervals for $f\sigma_8$ and $\sigma_v$ relative to those from the fiducial model in Fig. \ref{fig:sims_vv_mock1_systematics}.
The maximum likelihood when setting $k_{\rm{max}} = 0.10$\hmpc \ corresponds to $\chi^2 = 1055.09$ ($\chi^2/\text{dof} = 0.91$).
The maximum likelihood when setting $k_{\rm{max}} = 0.125$\hmpc \ corresponds to $\chi^2 = 1055.00$ ($\chi^2/\text{dof} = 0.91$).
We can see that changing the value of $k_{\rm{max}}$ has a negligible effect on the constraints from the logarithmic distance ratio auto-covariance.
This is supported by the $\chi^2$ values, which only vary on the order of $0.1$ between the three runs.
Consequently, we choose to keep our best estimate of $k_{\rm{max}}=0.15$\hmpc \ as our value for the fiducial model.

We also test how changing the value of the damping parameter $\sigma_u$ affects constraints.
We run the covariance model while setting $\sigma_u = 11.0, 15.0, 17.0$ \mpch \ and present the median values and 68\% credible intervals for $f\sigma_8$ and $\sigma_v$ relative to those from the fiducial model in Fig. \ref{fig:sims_vv_mock1_systematics}.
Again, the difference between our tested values is negligible, although we do see a slight trend in $f\sigma_8$, where lower values of $\sigma_u$ correspond to lower growth rates. 
This is consistent with the behaviour of the damping function, where lowering $\sigma_u$ results in less damping and a higher growth rate.
Similarly to the $k_{\rm{max}}$ test, the $\chi^2$ values only vary on the order of $0.1$.
Given that our best estimate of $\sigma_u$ provides a good fit, we choose to keep this as the value for our fiducial model.

Out of the two fixed model parameters, Fig. \ref{fig:sims_dd_mock1_systematics} shows us that $\sigma_u$ has the largest effect on $f\sigma_8$, but it is still minimal. 
$\sigma_v$ is unaffected by both the values of $k_{\rm{max}}$ and $\sigma_u$.
For $f\sigma_8$, the various tests all return median values that are well within 1$\sigma$ of our fiducial case.
We later use these tests to quantify the systematic uncertainty for our parameter estimates when working with the 6dFGS data in Section \ref{subsec:systematicuncert}.

\subsection{Complete Covariance} \label{subsec:sims_cov}
When running the full covariance model, we use the fiducial set-up that we established in the previous two sections: $b_{\rm{add}}\sigma_8$ included as a free parameter, $k_{\rm{max}} = 0.15$\hmpc \ for both auto-covariances and the cross-covariance, $\sigma_g = 3.0$\mpch \ for the galaxy overdensity auto-covariance and cross-covariance, and $\sigma_u = 13.0$\mpch \ for the logarithmic distance ratio auto-covariance and cross-covariance.
In addition to these, we have also introduced a new parameter for the cross-covariance, $\alpha_b$, which modifies the effective galaxy bias for the cross-covariance relative to that of the galaxy overdensity auto-covariance.

\begin{figure}
	\includegraphics[width=\columnwidth]{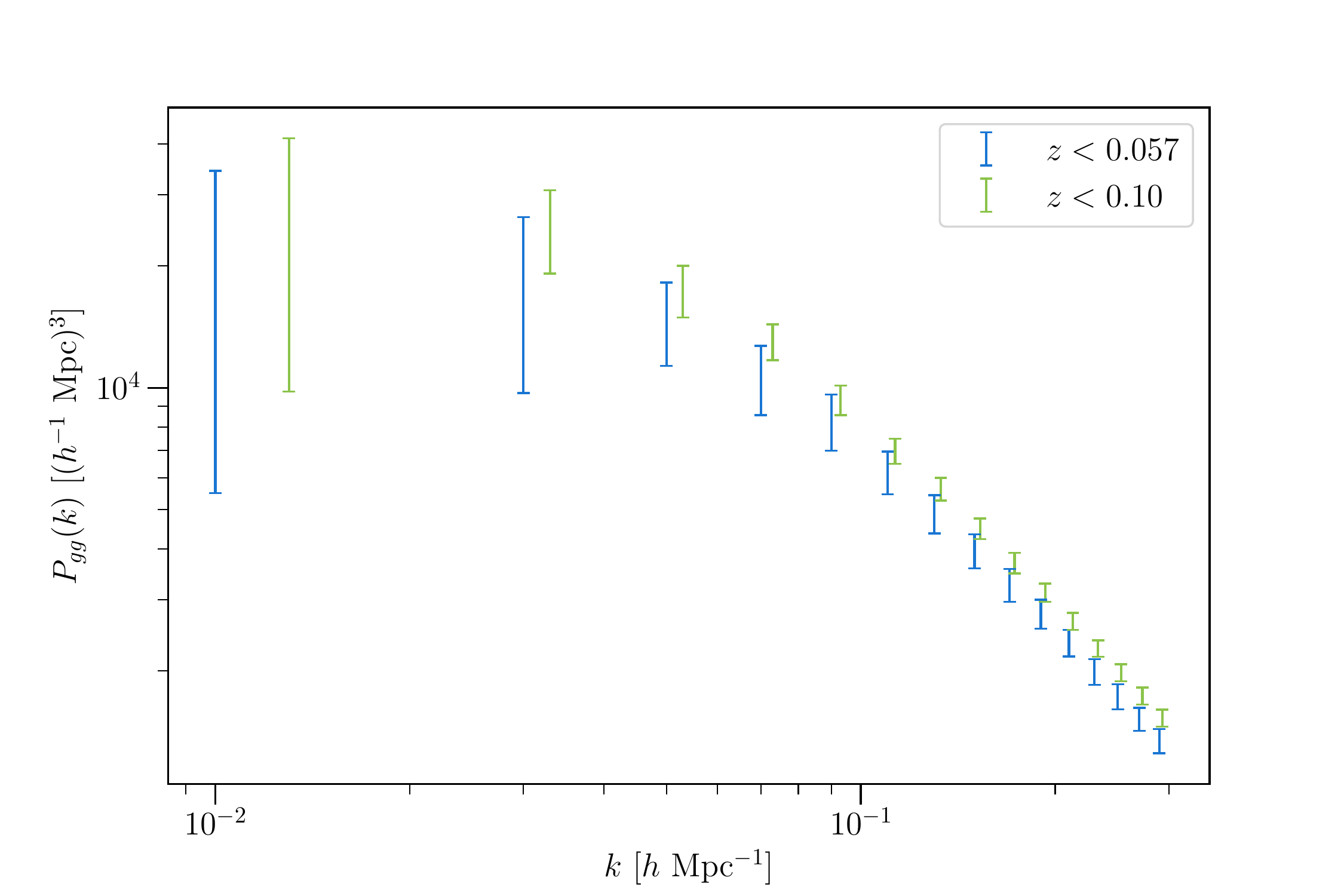}
	\caption{The mean and standard deviation of the galaxy-galaxy power spectrum from 600 mock catalogues, with uncertainties corresponding to those for one mock. The amplitude of the power spectrum for the $z<0.057$ sample is clearly lower than that for the $z<0.10$ sample, indicating that that the lower redshift sample has a lower effective bias. We note that the $z<0.10$ points have been shifted to the right by $\Delta k = 0.003$\hmpc \ for clarity.}
	\label{fig:sims_psredshiftdiff}
\end{figure}

Before proceeding, we establish that the effective bias of our sample is different for different redshift ranges.
We do this by estimating the galaxy-galaxy power spectrum for two redshift ranges, using the 600 mock catalogues developed by \cite{Carter2018} and discussed in Section \ref{subsec:sims}.
The redshift ranges correspond to the limits of our logarithmic distance ratio sample ($z<0.057$) and our galaxy overdensity sample ($z<0.1$).
We show the two estimated power spectra in Fig. \ref{fig:sims_psredshiftdiff}.

Nominally, we could estimate $\alpha_b^2$ directly by taking the ratio of the two estimated power spectra (recalling that the amplitude of the galaxy power spectrum is proportional to $b^2$).
The ratio between each pair of points implies $\alpha_b = 0.94\pm0.03$, where the uncertainty is scaled to a single mock.
However, we still choose to run the model for different values of $\alpha_b$ before selecting the fiducial value.
We do this for two key reasons.
Firstly, it is not trivial to estimate the redshift range that the logarithmic distance ratio sample (and hence the cross-covariance) is sensitive to.
\cite{Adams2017} showed that the cross-correlation is non-zero up to separations of at least $50$\mpch, depending on the orientation of the galaxy overdensity-logarithmic distance ratio pair (see fig. 9 and surrounding text in that work). 
This means that the effective bias for the cross-correlation is likely affected by overdensities beyond the boundary of the logarithmic distance ratio sample, a subtlety that the estimated value of $\alpha_b$ from the power spectra ratio does not account for.
Secondly, $\alpha_b$ may be sensitive to additional effects beyond the difference in effective bias.
For example, Eq. \ref{eq:modified_dv_alphab} and \ref{eq:modified_vd_alphab} show that $\alpha_b$ reduces the amplitude of the cross-covariance relative to the two auto-covariances (although we note that $\alpha_b$ only reduces the amplitude of the $\beta^0$ term and does not affect the $\beta^1$ term, see Eq. \ref{eq:dvcov_components}).
Additionally, it's possible that the relative weight of different regions towards the signal-to-noise differs between the cross-covariance and the galaxy overdensity auto-covariance.
Given these two factors, we determine that it is more appropriate to estimate the value of $\alpha_b$ by requiring recovery of the expected $f\sigma_8$ value when working with the mocks.

\begin{figure}
	\includegraphics[width=\columnwidth]{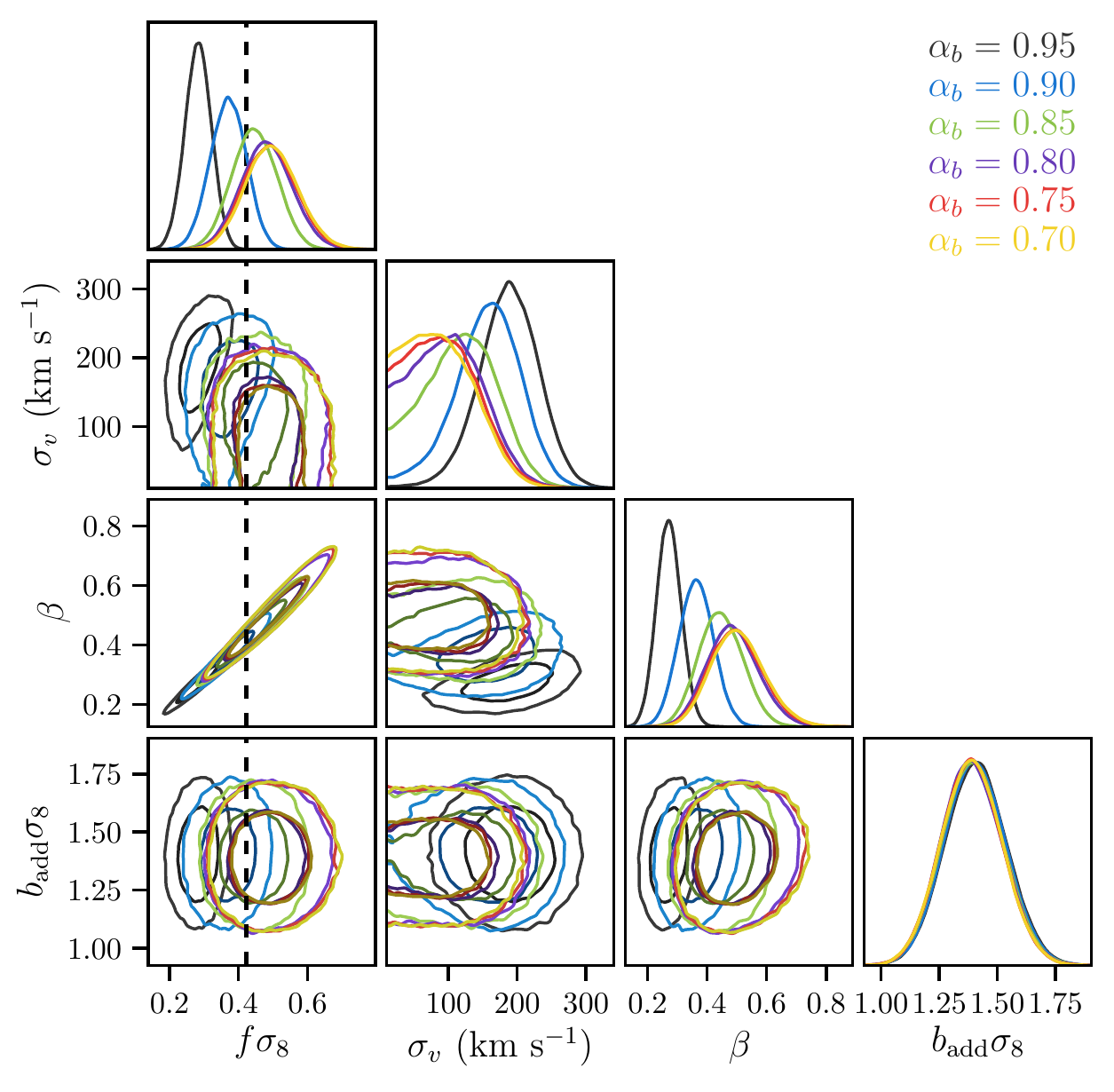}
	\caption{The posteriors of $f\sigma_8$, $\sigma_v$, $\beta$ and $b_{\rm{add}}\sigma_{8}$ for the complete covariance model fit to Mock 1 for different values of $\alpha_b$. The inner contour indicates the 68\% credible region and the outer contour indicates the 95\% credible region. The expected value for $f\sigma_8$ is shown by the dashed vertical line.}
	\label{fig:sims_cov_mock1_alphabvar}
\end{figure}

We start by running the full covariance model on our representative mock (Mock 1) with different values of $\alpha_b$. 
The lowest value we test is $\alpha_b = 0.70$; we consider lower values to be unphysical as they would translate to differences in the effective galaxy bias values that are implausible.
The posteriors are shown in Fig. \ref{fig:sims_cov_mock1_alphabvar}.

Fig. \ref{fig:sims_cov_mock1_alphabvar} indicates that for the representative mock, the growth rate is recovered at the 1$\sigma$ level for all values of $\alpha_b$ aside from $\alpha_b=0.95$. 
There is a clear systematic trend (although it is by no means linear) in that larger values of $\alpha_b$ correspond to lower values of $f\sigma_8$.
Additionally, we see that the posterior for $f\sigma_8$ widens as $\alpha_b$ becomes lower and that the posteriors appear to be converging.
This behaviour ties back to the fact that $\alpha_b$ reduces the amplitude of the cross-covariance model, causing it to contribute less to the overall covariance.
The increase in the value of $f\sigma_8$ and the widening of the posterior is consistent with the model favouring the auto-covariance information above the cross-covariance information. 
We note that the constraints from the low-$\alpha_b$ cases are still better than the galaxy overdensity auto-covariance only case, since the logarithmic distance ratio auto-covariance is also contributing to the constraints.
This interpretation also explains the behaviour of $\sigma_v$, which tends towards the logarithmic distance ratio auto-covariance constraint as $\alpha_b$ is lowered (see Fig. \ref{fig:sims_vv_mock1_fiducial} for comparison).
It's clear that the cross-covariance has little influence on $b_{\rm{add}}\sigma_8$, and consequently, the constraints are very close to those from the galaxy overdensity auto-covariance (see Fig. \ref{fig:sims_dd_mock1_fiducial}).

Given the systematic behaviour with $\alpha_b$, we choose to fix its value at $\alpha_b = 0.90$.
This maximises the amount of cross-covariance information used, while still recovering the growth rate of structure at the 1$\sigma$ level.
The maximum likelihood for this fit corresponds to $\chi^2=2615.90$ ($\chi^2/$dof = 0.94), which is reasonable.
We note that we account for the systematic behaviour from $\alpha_b$ by calculating a systematic error for our final 6dFGS growth rate constraint in Section \ref{subsec:systematicuncert}.

\begin{figure*}
	\includegraphics[width=0.75\textwidth]{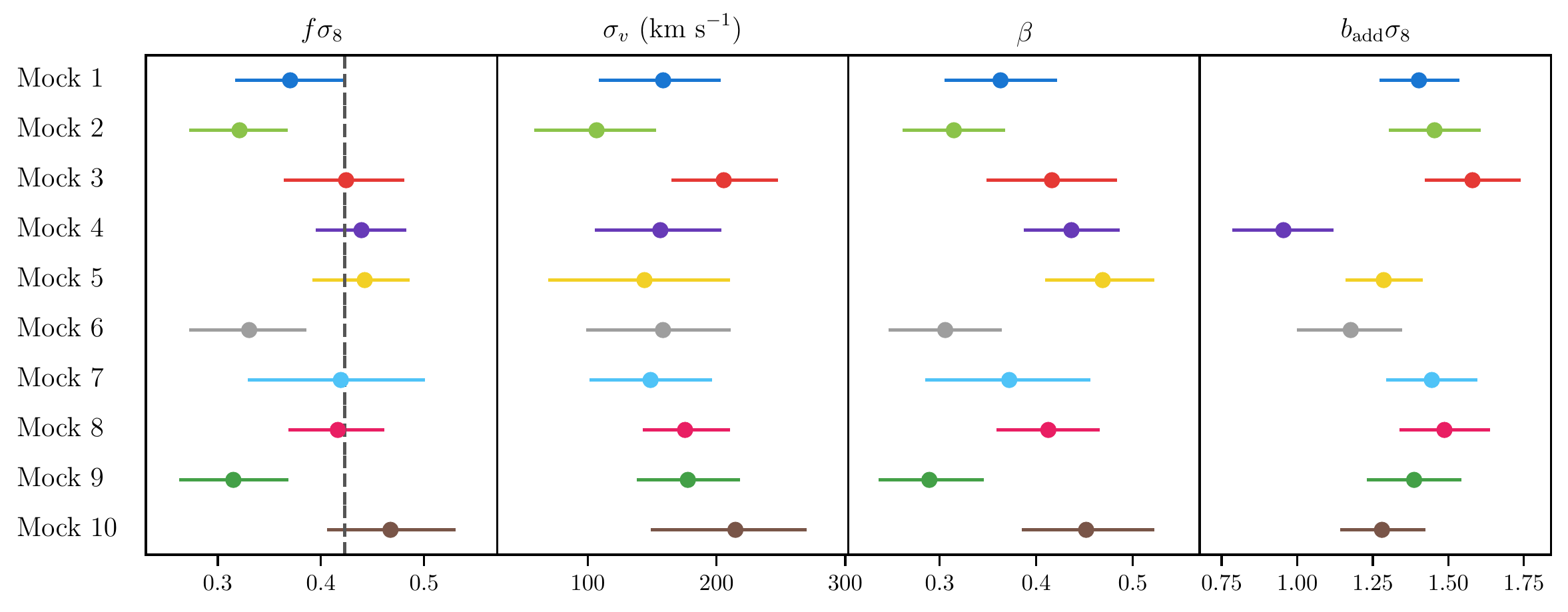}
	\caption{The median values and 68\% credible intervals of $f\sigma_8$, $\sigma_v$, $\beta$ and $b_{\rm{add}}\sigma_{8}$ for ten 6dFGS mocks when using the complete covariance with $\alpha_b = 0.90$. The expected value for $f\sigma_8$ is shown by the dashed vertical line.}
	\label{fig:sims_cov_allmocks_fiducial}
\end{figure*}

Finally, we show the median values and 68\% credible intervals for our free parameters when running the complete covariance model with $\alpha_b=0.90$ on all ten mocks in Fig. \ref{fig:sims_cov_allmocks_fiducial}.
The recovery of $f\sigma_8$ is reasonable across all ten mocks, validating our choice of $\alpha_b=0.90$ as the fiducial value for the remainder of the analysis.
Given that the mocks are independent, we also calculate the mean and error in the mean for the growth rate, finding $f\sigma_{8, \text{mean}} = 0.40 \pm 0.02$.
We note that the mean growth rate is consistent with the fiducial $f\sigma_8$ value at the 1$\sigma$ level.

\subsection{Summary of Fiducial Model Results} \label{subsec:chap4_simres_comparison}
We summarise the key constraints from running our fiducial model on our representative mock for the three covariance model cases in Table \ref{tab:6dfmocks_summary}.
\begin{table}
	\centering
	\caption{Median values (with 68\% credible intervals) of $f\sigma_8$, $\sigma_v$, $\beta$ and $b_{\rm{add}}\sigma_{8}$ for the galaxy overdensity auto-covariance, logarithmic distance ratio auto-covariance and complete covariance models for Mock 1.}
	\label{tab:6dfmocks_summary}
	\begin{tabular}{lcccc}
		\hline
		Model & $f\sigma_8$ & $\sigma_v$ (\kms) & $\beta$ & $b_{\rm{add}}\sigma_{8}$ \\ 
		\hline
		$\bm{\mathbfss{C}}_{\delta_g \delta_g, \rm{RSD}}^{\rm err}$ & $0.47\pm 0.13$ & -- & $0.47^{+0.16}_{-0.14}$ & $1.38\pm 0.13$ \\ 
		$\bm{\mathbfss{C}}_{\eta \eta, \rm{RSD}}^{\rm err}$ & $0.43^{+0.11}_{-0.10}$ & $84^{+57}_{-49}$ & -- & -- \\ 
		$\bm{\mathbfss{C}}$ & $0.370^{+0.053}_{-0.052}$ & $159^{+45}_{-50}$ & $0.363\pm 0.058$ & $1.40\pm 0.13$\\
		\hline
	\end{tabular}
\end{table}
It is clear that the uncertainty in the growth rate of structure has reduced significantly when using the complete covariance in comparison to using either of the auto-covariances alone: we see a 60\% improvement in the uncertainty from the galaxy overdensity auto-covariance and a 50\% improvement in the uncertainty from logarithmic distance ratio auto-covariance.
For $\beta$ the uncertainty improvement is 61\% when going from the galaxy overdensity auto-covariance to the complete covariance.
Given the sophistication of the mocks, we expect to see similar improvements when applying our fiducial model to the data in the next section.

While we believe that the reduction in uncertainty can be entirely attributed to the introduction of the cross-covariance, we note that tighter uncertainties can be a symptom of underlying tension in the model or data sets.
Given the consistency of the growth rate constraints and that the complete covariance model shows a reasonable $\chi^2$/dof (0.94), we do not believe that tension is impacting the uncertainty reduction.

\section{Data results} \label{sec:6dfgsresults}
\begin{figure*}
	\includegraphics[width=0.7\textwidth]{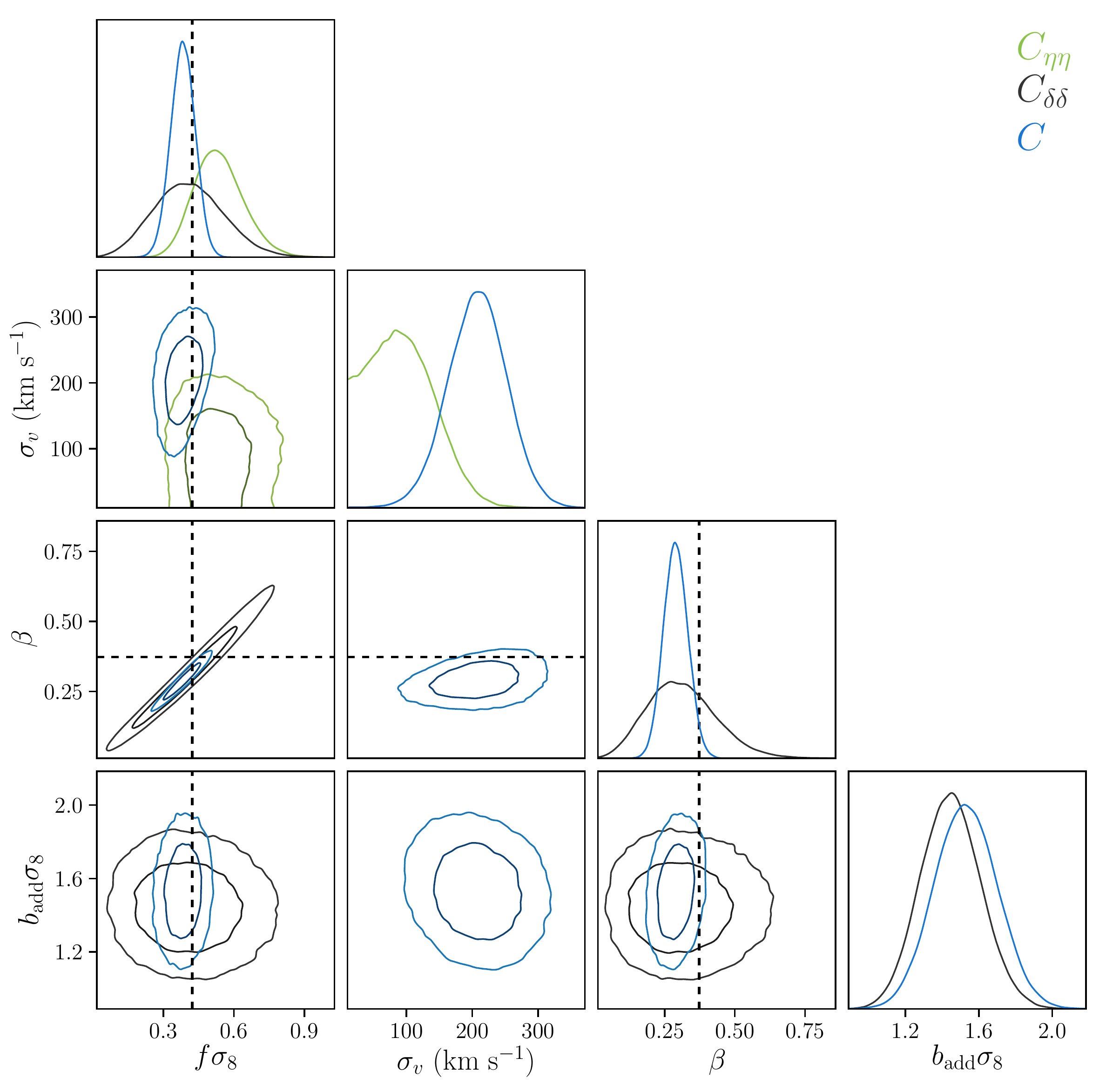}
	\caption{The posteriors of $f\sigma_8$, $\beta$, $b_{\rm{add}}\sigma_{8}$ and $\sigma_v$ for 6dFGS when using the galaxy overdensity auto-covariance ($\mathbfss{C}_{\delta\delta}$), the logarithmic distance ratio auto-covariance ($\mathbfss{C}_{\eta\eta}$), and the complete covariance ($\mathbfss{C}$). The results for $f\sigma_8$ and $\beta$ from \citet{Beutler2012} are indicated by the dashed  lines.}
	\label{fig:data_cov_comparison}
\end{figure*}

After comprehensive testing on the 6dFGS mock catalogues, we run the galaxy overdensity auto-covariance, logarithmic distance auto-covariance and complete covariance models on the 6dFGS dataset.
The models use the fiducial parameter values from the previous section: $k_{\rm{max}} = 0.15$\hmpc, $\sigma_g = 3.0$\mpch, $\sigma_u = 13.0$\mpch, $r_g = 1.0$, and $\alpha_b = 0.90$.
Now that we are working with real data, we choose the fiducial cosmology to be that from \cite{PlanckCollaboration2015a}, which is summarised in Table \ref{tab:powerspectrummodels}.
Given that we are working with a complete RSD model for our overdensity data, we directly compare our results to those from \cite{Beutler2012}, who found $f\sigma_8 = 0.423 \pm 0.055$ and $\beta = 0.373 \pm 0.054$ at an effective redshift of $z_{\rm{eff}} = 0.067$.
These are the most precise measurements of $f\sigma_8$ and $\beta$ available for 6dFGS, so serve as a useful point of comparison.
We present the posteriors of our free parameters in Fig. \ref{fig:data_cov_comparison}, the corresponding median constraints (with 68\% confidence intervals) in Table \ref{tab:6dfgs_med} and maximum likelihood values (with the corresponding $\chi^2$ values) in Table \ref{tab:6dfgs_max}.
For the galaxy overdensity auto-covariance we find $f\sigma_{8} = 0.41^{+0.15}_{-0.14}$, for the logarithmic distance ratio auto-covariance we find $f\sigma_{8} = 0.53^{+0.11}_{-0.10}$, and for the complete covariance we find $f\sigma_{8} = 0.384\pm{0.052}$.

\begin{table}
	\centering
	\caption{Median values (with 68\% credible intervals) of $f\sigma_8$, $\sigma_v$, $\beta$ and $b_{\rm{add}}\sigma_{8}$ for 6dFGS using the galaxy overdensity auto-covariance ($\mathbfss{C}_{\delta\delta}$), the logarithmic distance ratio auto-covariance ($\mathbfss{C}_{\eta\eta}$), and the complete covariance ($\mathbfss{C}$).}
	\label{tab:6dfgs_med}
	\begin{tabular}{lcccc}
		\hline
		Model & $f\sigma_8$ & $\sigma_v$ (\kms) & $\beta$ & $b_{\rm{add}}\sigma_{8}$ \\ 
		\hline
		$\mathbfss{C}_{\delta\delta}$ & $0.41^{+0.15}_{-0.14}$ & -- & $0.30^{+0.13}_{-0.11}$ & $1.45^{+0.17}_{-0.16}$ \\ 
		$\mathbfss{C}_{\eta\eta}$ & $0.53^{+0.11}_{-0.10}$ & $90^{+54}_{-50}$ & -- & -- \\ 
		$\mathbfss{C}$ & $0.384\pm 0.052$ & $208^{+44}_{-45}$ & $0.289^{+0.044}_{-0.043}$ & $1.53\pm 0.17$ \\ 
		\hline
	\end{tabular}
\end{table}

\begin{table}
	\centering
	\caption{Maximum likelihood values of $f\sigma_8$, $\sigma_v$, $\beta$ and $b_{\rm{add}}\sigma_{8}$ for 6dFGS using the galaxy overdensity auto-covariance ($\mathbfss{C}_{\delta\delta}$), the logarithmic distance ratio auto-covariance ($\mathbfss{C}_{\eta\eta}$), and the complete covariance ($\mathbfss{C}$). We also include the $\chi^2$ and $\chi^2$/dof statistic for the maximum likelihood values}
	\label{tab:6dfgs_max}
	\begin{tabular}{lcccccc}
		\hline
		Model & $f\sigma_8$ & $\sigma_v$ (\kms) & $\beta$ & $b_{\rm{add}}\sigma_{8}$ & $\chi^2$ & $\chi^2$/dof \\ 
		\hline
		$\mathbfss{C}_{\delta\delta}$ & $0.38$ & -- & $0.27$ & $1.45$ & 1774.45 & 1.09 \\ 
		$\mathbfss{C}_{\eta\eta}$ & $0.52$ & 84 & -- & -- & 847.69 & 0.94 \\ 
		$\mathbfss{C}$ & $0.380$ & $208$ & $0.286$ & $1.52$ & 2610.42 & 1.03 \\ 
		\hline
	\end{tabular}
\end{table}

We find that our measurements of $f\sigma_8$ and $\beta$ for the three covariance analyses are self-consistent.
We have hence demonstrated that peculiar velocities and redshift-space distortions provide consistent measurements of the growth rate of structure for the same galaxy survey and modelling framework.
Given that the two probes constrain the growth rate on different physical scales (peculiar velocities are sensitive to larger scales than RSD), the complete covariance analysis may be a promising way to test modified gravity models, which is a promising avenue for future work.
We also note that our value of $f\sigma_8$ is consistent at close to the 1$\sigma$ level with the Planck 2015 + GR prediction of $f\sigma_8=0.446$ at redshift $z=0$ and at the 1$\sigma$ level with the Planck 2018 + GR prediction of $f\sigma_8 = 0.430$ at redshift $z=0$.

We calculate that the percentage uncertainties in $f\sigma_8$ are 35\% for the galaxy overdensity auto-covariance, 20\% for the logarithmic distance ratio auto-covariance and 14\% for the complete covariance.
Most importantly, we see a 64\% reduction in the uncertainty when going from the galaxy overdensity auto-covariance to the complete covariance, and a 50\% reduction when going from the logarithmic distance ratio auto-covariance to the complete covariance.
The improvement in going from the galaxy overdensity auto-covariance to the complete covariance is most notable in the $f\sigma_8$-$\beta$ contour of Fig. \ref{fig:data_cov_comparison}, where the 2$\sigma$ contour from the complete covariance sits well inside the 1$\sigma$ contour from the auto-covariance.

It's also worth highlighting that the consistent $f\sigma_8$-$\beta$ slope between the galaxy overdensity auto-covariance and the complete covariance indicates that the two models prefer similar effective galaxy bias values.
We note that this would not be the case without an appropriate value for $\alpha_b$, which allows the cross-covariance to be parametrized in terms of the galaxy overdensity sample's effective bias.

\subsection{Systematics} \label{subsec:systematicuncert}
Given the increased precision of our constraint on the growth rate of structure, it is important to investigate how robust our result is to various systematics.
This includes the fixed parameters of our covariance model ($k_{\rm{max}}$, $\sigma_g$, $\sigma_u$ and $\alpha_b$), as well as the underlying cosmological parameters which inform the power spectrum models.
Note that we do not investigate the systematic effects of changing $r_g$, as we chose to leave this as a fixed parameter corresponding to the assumption of the linear bias model (see Section \ref{subsec:sims_ddcov}).
Future work could potentially vary this parameter, although we note it is highly degenerate with the growth rate of structure.

\subsubsection{Sensitivity to Fixed Parameters}
For the fixed parameters, we're able to estimate a systematic error contribution by 
varying the values of the fixed parameters and re-running the model.
We assume that each systematic is independent, allowing us to vary a single parameter while holding the others fixed at their fiducial values.
For each systematic $s \in (k_{\rm{max}}, \sigma_g, \sigma_u, \alpha_b)$, the systematic variance in parameter $\phi$ is
\begin{align}
\sigma_s^2 = \left( \frac{\partial \phi}{\partial s} \right)^2 (\Delta s)^2,
\end{align}
where we approximate the partial derivative using the central finite difference method:
\begin{align}
\frac{\partial \phi}{\partial s} \approx \frac{\phi(s+\Delta s) - \phi(s-\Delta s)}{2\Delta s}.
\end{align}
We note that $\sigma_s^2$ are the diagonal elements of the full systematic covariance \citep[e.g. eq. C4 in][]{Zhang2017}.
We then give the total systematic error as the sum in quadrature of each systematic:
\begin{align}
\sigma_{\rm{sys}} = \sqrt{\sum_{i=s} \sigma_i^2}.
\end{align}

We note that the size of the systematic variance will be affected by the step size $\Delta s$.
Consequently, we mostly use the same step sizes that we used when testing each fixed parameter throughout the simulation analysis in Section \ref{sec:simresults}, which were chosen to encompass reasonable values for the fixed parameters. 
The only exception is in the case of $\alpha_b$, where we choose a smaller step size of $\Delta\alpha_b = 0.025$, as the step size of $\Delta\alpha_b = 0.05$ gives posteriors that recover the growth rate outside the $1\sigma$ level (see Fig. \ref{fig:sims_cov_mock1_alphabvar}).
The systematic standard deviation values are given in Table \ref{tab:6dfgs_sysvar} for $f\sigma_8$ and $\beta$, noting that we don't provide systematic standard deviation estimates for our two nuisance parameters $\sigma_v$ and $b_{\rm{add}}\sigma_8$ since they are already marginalised over in the model fits. 
\begin{table}
	\centering
	\caption{The systematic standard deviation contributions to $f\sigma_8$ and $\beta$ for each fixed parameter.}
	\label{tab:6dfgs_sysvar}
	\begin{tabular}{lcccc}
		\hline
		Parameter & $\sigma_{k_{\rm{max}}}$ & $\sigma_{\sigma_g}$ & $\sigma_{\sigma_u}$ & $\sigma_{\alpha_b}$  \\ 
		\hline
		$f\sigma_8$ & $1.69\times10^{-3}$ & $2.84\times10^{-3}$ & $1.09\times10^{-3}$ & $6.06\times10^{-2}$ \\ 
		$\beta$ & $3.30\times10^{-3}$ & $8.97\times10^{-4}$ & $1.08\times10^{-3}$ & $4.84\times10^{-2}$ \\ 
		\hline
	\end{tabular}
\end{table}
It's clear that $\alpha_b$ is the dominant systematic for both $f\sigma_8$ and $\beta$, with a systematic standard deviation that is at least an order of magnitude larger than any of the other fixed parameters.

From this analysis, our final constraint (using the full covariance model) is $f\sigma_8 = 0.384 \pm 0.052 \rm{(stat)} \pm 0.061 \rm{(sys)}$ for the growth rate of structure, and $\beta = 0.289^{+0.044}_{-0.043} \rm{(stat)} \pm 0.049 \rm{(sys)}$ for the redshift-space distortion parameter.
Currently, the systematic error for each parameter is greater than the corresponding statistical uncertainty, which is driven by the behaviour of $\alpha_b$, specifically, its large degeneracy with $f\sigma_8$. 
Given that our introduction of $\alpha_b$ is a relatively simple method for accounting for the difference in effective bias across our samples, we believe that this systematic could be reduced or mitigated in future work, and suggest some avenues for this in Section \ref{subsec:futurework}.

\subsubsection{Sensitivity to Cosmological Parameters}
Our method is affected by the cosmological parameter values (those listed in Table \ref{tab:powerspectrummodels}) in two key ways.
Firstly, through the transformation of the observed coordinates (RA, dec, $z$) to Cartesian coordinates ($x$, $y$, $z$) in configuration space, which is required for our covariance model.
Secondly, the cosmological parameters influence the shape of all three model power spectra $P_{mm}$, $P_{\theta\theta}$ and $P_{m\theta}$.
We note that the second point is more important, since the transformations from observed to Cartesian coordinates are independent of all the cosmological parameters to first order, being at low redshift and with distances measured in \mpch \ units.

\begin{figure*}
	\includegraphics[width=0.7\textwidth]{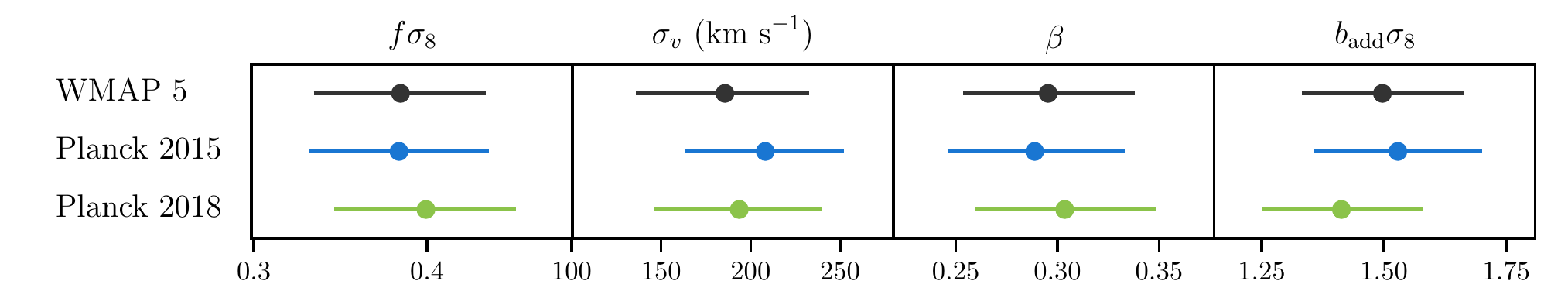}
	\caption{The median values and 68\% credible intervals of $f\sigma_8$, $\sigma_v$, $\beta$ and $b_{\rm{add}}\sigma_{8}$ for 6dFGS when using three sets of cosmological parameters, as listed in Table \ref{tab:powerspectrummodels}.}
	\label{fig:data_cov_cosmologysys}
\end{figure*}

To test how sensitive our $f\sigma_8$ constraint is to the choice of the cosmological parameter values, we use the values from the three CMB analyses listed in Table \ref{tab:powerspectrummodels}: the WMAP Year-5 results \citep{Komatsu2009}, Planck 2015 results \citep[our fiducial model;][]{PlanckCollaboration2015a} and the Planck 2018 results \citep{PlanckCollaboration2018}.
We repeat our analysis for the complete covariance, including the data transformation, for the two additional cosmological parameter sets, and present the median values and 68\% credible intervals for $f\sigma_8$ ,$\sigma_v$, $\beta$ and $b_{\rm{add}}\sigma_8$ for all three parameter sets in Fig. \ref{fig:data_cov_cosmologysys}.
It's clear from the figure that the choice of cosmological parameters has little effect on $f\sigma_8$.

\subsection{Comparison to Existing Literature}
\subsubsection{Covariance Analysis} \label{subsubsec:comp_completecov}
Using a combined auto- and cross- covariance model without redshift-space distortions (RSD), \cite{Adams2017} found $f\sigma_8 = 0.424^{+0.065}_{-0.062}$ and $\beta = 0.300^{+0.048}_{-0.046}$, which we compare to the constraints from this analysis of $f\sigma_8 = 0.384 \pm 0.052 \rm{(stat)} \pm 0.061 \rm{(sys)}$ and $\beta = 0.289^{+0.044}_{-0.043} \rm{(stat)} \pm 0.049 \rm{(sys)}$.
For $f\sigma_8$, the statistical uncertainty is reduced by $18\%$ and our result is consistent at the 1$\sigma$ level.
For $\beta$, the statistical uncertainty is reduced by $7.4\%$ and our result is consistent at the 1$\sigma$ level.
We believe that the improvement in the statistical uncertainty comes from two sources: the improved covariance model, which now utilises the growth rate information present in RSD, and the larger galaxy overdensity sample used in this analysis.
The improvement from the larger sample is due to two factors: there are more covariance entries because we've extended to higher redshift ($N_{\delta_g} = 1633$ cells compared to $N_{\delta_g} = 1036$ cells) and a larger number of galaxies per cell at low redshift (which comes from no longer having the volume-limited sample), reducing the shot noise.
We suspect that the overall improvement in the statistical uncertainty may also be slightly limited by the fact that we used a larger gridding scale, smoothing over more of the small-scale information.
We show a visual comparison of the covariance constraints in Fig. \ref{fig:literaturecomp}.

\subsubsection{Multi-Tracer Approaches}
In this study, we have demonstrated the benefits of using the shared information from multiple tracers, which supports the results of theoretical studies \citep[e.g.][]{McDonald2009,Gil-Marin2010,Bernstein2011,Abramo2013}. 
We compare the improvements we see in the statistical uncertainty to those seen by \cite{Blake2013}, who presented the first multi-tracer approach applied to galaxy overdensity data from the Galaxy And Mass Assembly (GAMA) survey.
In that study, they used two different galaxy overdensity samples with different galaxy bias values, comparing the power spectra of these samples to models, including the cross-power spectrum.
Depending on the sample, they found a 10-20\% improvement in their constraints of the growth rate when utilising the cross-power spectrum.
In our analysis, we see significantly better improvements: 50\% improvement compared to the logarithmic distance ratio only sample, and 64\% improvement compared to the galaxy overdensity only sample.
We believe that this is due, in part, to the strong independent constraints that the logarithmic distance ratio places on the growth rate. 
When coupled with the fact that the two samples (and their cross-correlation) constrain the growth rate in different ways, we believe that this explains our larger improvement on the growth rate constraint compared to the analysis from \cite{Blake2013}, which only utilised RSD information.

\subsubsection{Forecasts for the 6dFGS Cross-Covariance Analysis}
We also compare our relative constraints to those forecast for 6dFGS from two studies that use the Fisher matrix formalism: \cite{Koda2014} and \cite{Howlett2017}.
Both studies use the same RSD model as we do and forecast the relative constraints that can be obtained on $f\sigma_8$ for various samples of 6dFGS. 

\cite{Koda2014} use 6dFGSv as the basis for both their galaxy overdensity and logarithmic distance ratio sample, and model the various covariances out to $k_{\rm{max}} = 0.1$\hmpc.
They forecast a 25\% constraint on $f\sigma_8$ for the logarithmic distance ratio auto-covariance and 15\% when using the complete covariance. 
For $\beta$, they forecast a 16\% constraint when using the complete covariance.
We find a relative constraint on $f\sigma_8$ of 20\% for the logarithmic distance ratio auto-covariance and 14\% for the complete covariance, and a relative constraint on $\beta$ of 15\% for the complete covariance.
In this analysis, we find a constraint from the logarithmic distance ratio auto-covariance that is better than forecast, and note that this was also the case in \cite{Adams2017} and \cite{Johnson2014}.
This could be due to differences between the assumptions that went into the forecasting and our own analysis.
We should expect to do better than \cite{Koda2014} because our galaxy overdensity sample goes to a higher redshift ($z=0.1$ compared to $z=0.057$), which we find to be the case, but only by a single percentage point.

\cite{Howlett2017} used the complete 6dFGSv sample as the basis for their logarithmic distance ratio sample and close to the complete 6dFGSz sample as the basis for their galaxy overdensity sample (the upper limit on the redshift for their sample is $z=0.2$).
Like \cite{Koda2014}, they model the various covariances out to $k_{\rm{max}} = 0.1$\hmpc.
They forecast a 25.1\% constraint on $f\sigma_8$ for the logarithmic distance ratio auto-covariance and 11.2\% when using the complete covariance. 
For $\beta$, they forecast a 12.3\% constraint when using the complete covariance.
The similarity of their logarithmic distance ratio auto-covariance constraint to that from \cite{Koda2014} is consistent with the fact that they used very similar samples.
The forecast constraints are better than our statistical uncertainties, which is unsurprising given they used a larger galaxy overdensity sample.

As a final point, we note that both of these analyses assume that the galaxy overdensity auto-correlation and cross-correlation are responding to the same effective bias. 
At this stage, it is unclear what effect this might have on the precision of forecasting, since we found that the value of the relative effective bias (parametrized by $\alpha_b$ in our study) has a significant effect on the posterior of $f\sigma_8$.

\subsubsection{6dFGS Redshift-Space Distortion and Velocity Results}
One of the clear advantages of our approach is that it provides a new method to constrain the growth rate of structure.
Consequently, it is informative to compare our results to those from other analyses of 6dFGS.
Several RSD analyses have been performed using 6dFGS: \cite{Beutler2012} presented a traditional RSD analysis, \cite{Achitouv2017} looked at RSD in the void-galaxy cross-correlation, and \cite{Blake2018} presented a Fourier-space analysis using the RSD power spectrum multipoles. 
In addition to these, \cite{Johnson2014} and \cite{Huterer2017} both presented logarithmic distance ratio auto-covariance analyses using the 6dFGSv sample.
Finally, \cite{Qin2019} applied a new estimator of the redshift-space density and momentum power spectra to redshifts and peculiar velocities from 6dFGSv.

Our galaxy overdensity auto-covariance analysis is most comparable to the results from \cite{Beutler2012}. 
However, there are some minor differences that should be kept in mind.
The galaxy overdensity sample used by \cite{Beutler2012} has a slightly lower magnitude cut ($K \leq 12.75$ compared to $K \leq 12.9$), and uses galaxies from a larger redshift range ($z \lesssim 0.2$), yielding 81,971 galaxies compared to the 70,467 galaxies used in our analysis.
We also note that the \cite{Beutler2012} analysis employs the Feldman-Kaiser-Peacock \citep[FKP;][]{Feldman1994} weighting scheme to improve their statistical constraints, where we do not.
By measuring the correlation function, they found $f\sigma_8 = 0.423 \pm 0.055$ (13\% relative uncertainty) and $\beta = 0.373 \pm 0.054$ (14\% relative uncertainty).
These results are significantly better (although still consistent at the 1$\sigma$ level) than our galaxy overdensity auto-covariance results of $f\sigma_8 = 0.41 ^{+0.15}_{-0.14}$ and $\beta = 0.30 ^{+0.13}_{-0.11}$.
There are several factors that could explain this: \cite{Beutler2012} use a higher redshift sample, FKP weighting, and have access to smaller-scale information, which we lose by smoothing our model after gridding.
We note that in terms of statistical uncertainties, our complete covariance constraints of $f\sigma_8 = 0.384 \pm 0.052$ and $\beta = 0.289^{+0.044}_{-0.043}$ are better than those from \cite{Beutler2012}, although this advantage is lost when considering the combined statistical and systematic uncertainty.

As in this work, the analysis by \cite{Achitouv2017} also uses the \cite{Beutler2012} galaxy overdensity sample as a starting point. 
For their void-galaxy cross-correlation analysis, they take a volume-limited sample out to redshift $z=0.05$, similar to the sample selection we made in our first analysis, and implement FKP weighting.
They find $f\sigma_8 = 0.39 \pm 0.11$ when fitting to the void-galaxy cross-correlation function, which is consistent with both our galaxy overdensity auto-covariance and complete covariance results at the 1$\sigma$ level.
Even with the lower redshift sample, this work provides a tighter constraint than our galaxy overdensity auto-covariance. 
We expect that the same factors that we highlighted when comparing to \cite{Beutler2012} are at play, especially the fact that the correlation fitting method may be accessing information on smaller scales.

\cite{Blake2018} presented an RSD analysis that fits to the power spectrum multipoles rather than the correlation function, making it the first Fourier-space analysis of RSD for 6dFGS.
We note that they used the same sample as us: the \cite{Beutler2012} sample out to redshift $z=0.1$, and they implement FKP weighting.
They find $f\sigma_8 = 0.38 \pm 0.12$, which is consistent with both our galaxy overdensity auto-covariance and complete covariance results at the 1$\sigma$ level.
In their analysis, they highlighted that their larger statistical uncertainty  \citep[relative to the standard correlation function analysis from][]{Beutler2012} was likely due to the correlation function analysis accessing smaller scale information than was available in the multipoles analysis.
This is consistent with the interpretation of our own results, and this coupled with the fact that \cite{Blake2018} also used FKP weighting could explain why our statistical uncertainty is slightly larger than theirs but more than double that from \cite{Beutler2012}. 

Our method has been largely informed by that of \cite{Johnson2014}, who effectively presented an logarithmic distance ratio auto-covariance analysis of 6dFGSv. 
They found $f\sigma_8 = 0.428 ^{+0.079}_{-0.068}$, which is consistent with our both our logarithmic distance ratio auto-covariance and complete covariance results at the 1$\sigma$ level.
We note that their constraint is better than our logarithmic distance ratio auto-covariance constraint of $f\sigma_8 = 0.53^{+0.11}_{-0.10}$.
We suspect that this may be due to the fact that \cite{Johnson2014} used a gridding scale of 10\mpch \ where we used 20\mpch. 
This would result in more covariance entries and potentially lower the statistical uncertainty.

\cite{Huterer2017} performed a very similar analysis to \cite{Johnson2014} using 6dFGSv, but did not grid their sample. 
They found $f\sigma_8 = 0.481^{+0.067}_{-0.064}$, which is again consistent with our logarithmic distance ratio auto-covariance and complete covariance results at the 1$\sigma$ level.
We note that the lower absolute statistical uncertainty relative to \cite{Johnson2014} could be to do with the number of entries in the covariance matrix.

Finally, \cite{Qin2019} applied an estimator of the redshift-space density and momentum power spectra to 6dFGSv, constraining the growth rate of structure by comparing their estimated power spectra to modelled power spectra. 
This is similar to our work in that it utilises both redshifts and peculiar velocities, and while the methods are different, we consider them to be highly complementary.
They found $f\sigma_8 = 0.451^{+0.108}_{-0.092}$, which is consistent with both of our auto-covariance results and the complete covariance result at the 1$\sigma$ level.

We show a visual comparison of our $f\sigma_8$ constraints to these existing 6dFGS constraints in Fig. \ref{fig:literaturecomp}.

\subsubsection{Density-Velocity Comparison Approaches}
Density-velocity comparison approaches also take advantage of the shared information between the galaxy overdensity and peculiar velocity fields, albeit in a different way to our method. 
These methods use gravitational instability theory to relate the galaxy overdensity field to the peculiar velocity field through
\begin{align}
\bm{v}_p(\bm{r}) = \frac{H_0\beta}{4\pi} \int d^3\bm{r'} \frac{\delta_g(\bm{r'})(\bm{r'}-\bm{r})}{|\bm{r'} - \bm{r}|^3}.
\end{align}
This relation can be used to predict the peculiar velocity field from the observed galaxy overdensity field; comparing the modelled field to the observed field then allows one to constrain $\beta$.
The growth rate can be extracted from this if one has an estimate of the galaxy bias for the sample.

The three studies we compare our results to are \cite{Pike2005}, \cite{Davis2011} and \cite{Carrick2015}.
We note that all three use variations of the 2-Micron All-Sky Survey (2MASS) for the galaxy overdensity sample and variations of the Spiral Field I-Band survey for the velocity sample.
\cite{Pike2005} found $f\sigma_8 = 0.44 \pm 0.06$, \cite{Davis2011} found $f\sigma_8 = 0.32 \pm 0.04$ and \cite{Carrick2015} found $f\sigma_8 = 0.427 \pm 0.027$. 
Our complete covariance constraints are consistent with each of these at the 1$\sigma$ level, both with and without the systematic error, and our statistical uncertainty is between that of \cite{Pike2005} and \cite{Davis2011}.
We show a visual comparison of our complete covariance constraint for $f\sigma_8$ to these existing velocity-velocity constraints in Fig. \ref{fig:literaturecomp}.

\subsubsection{Cross-Correlation Only Analysis}
\begin{figure}
	\includegraphics[width=\columnwidth]{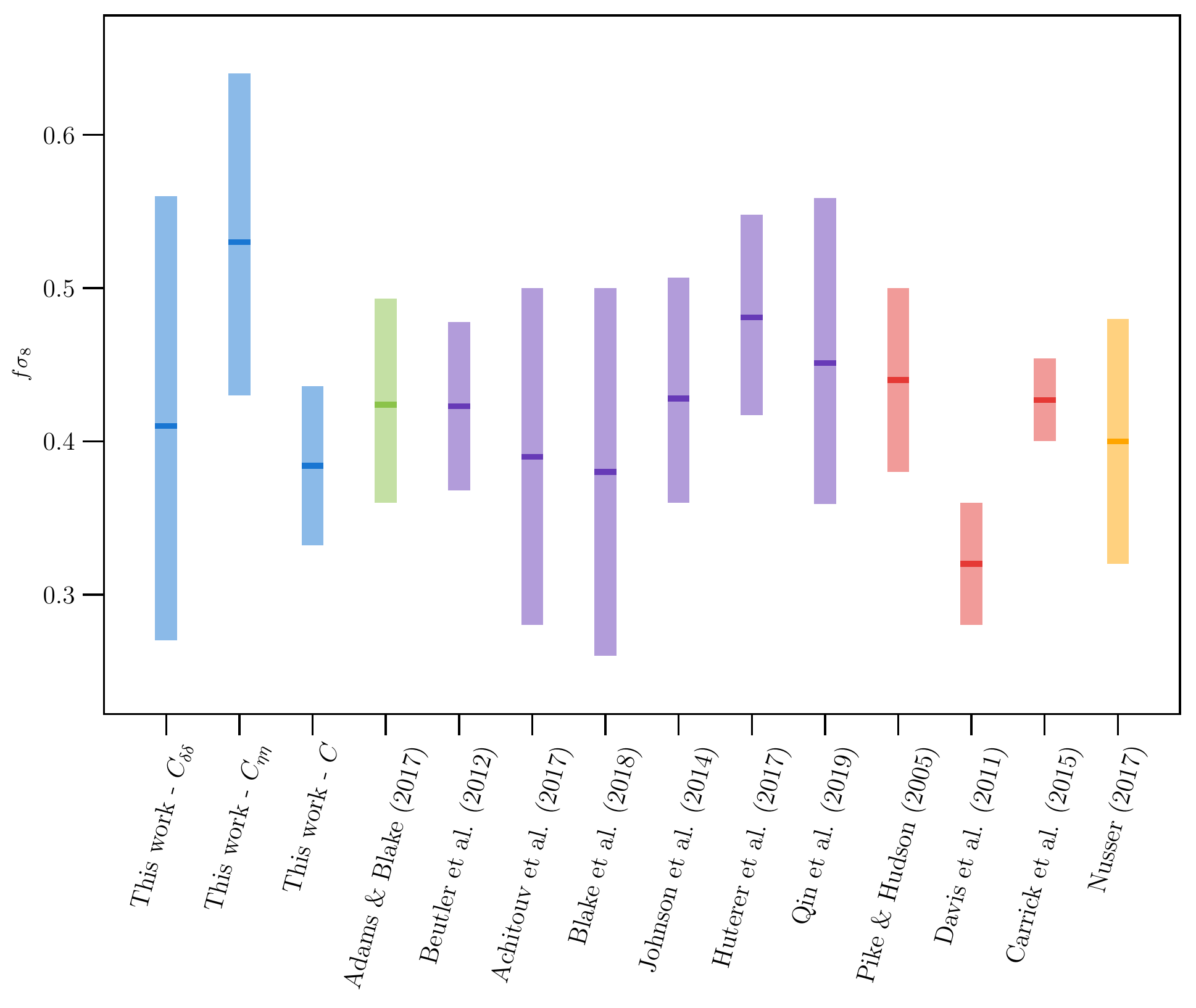}
	\caption{Median values (solid bar) and 68\% credible interval (shaded region) of $f\sigma_8$ for this work (shown in blue) and \citet{Adams2017} (shown in light green). Other works utilising 6dFGS are shown in purple, velocity-velocity comparisons are shown in red and the cross-correlation only analysis is shown in orange.}
	\label{fig:literaturecomp}
\end{figure}

Finally, we compare our constraint of $f\sigma_8$ to that from \cite{Nusser2017}, who presented a fit to the cross-correlation function for galaxy overdensities (from 2MASS) and peculiar velocities (from the \textit{cosmicflows-3} catalogue).
This is more similar to the analysis by \cite{Achitouv2017} than our analysis in that they model the cross-correlation as a function of separation, similar to how \cite{Achitouv2017} modelled the cross-correlation function between galaxies and voids.
They found $f\sigma_8 = 0.40 \pm 0.08$, which is consistent with our complete covariance constraint at the 1$\sigma$ level.
The construction of this method means they only utilise the equivalent of our cross-covariance, rather than the complete covariance. This explains why we see tighter statistical uncertainties. 
We show a visual comparison of our complete covariance constraint for $f\sigma_8$ to this constraint in Fig. \ref{fig:literaturecomp}.

\subsection{Future Work} \label{subsec:futurework}
We now highlight possible improvements and new research opportunities that arise from this work.

In Section \ref{subsubsec:alphab} we introduced $\alpha_b$ to parametrize the difference in the effective galaxy bias of the galaxy overdensity auto-correlation and cross-correlation.
While we found that this parameter was sufficient to recover the growth rate of structure in the mocks, it was the dominant source of systematic uncertainty in our final constraints. 
This can be linked to the fact that $\alpha_b$ represents more than just the difference between effective bias values; it directly influences the amplitude of the cross-covariance, such that a lower value of $\alpha_b$ may signify that the cross-correlation between peculiar velocities and galaxy overdensities is weaker than anticipated by our model.
We discovered this behaviour when fitting our model with different values of $\alpha_b$, noting that the posteriors on $f\sigma_8$ appeared to converge for increasingly small values of $\alpha_b$ (see Fig. \ref{fig:sims_cov_mock1_alphabvar}).

Noting that the difference in effective bias arises from being unable to use a volume-limited sample, we can turn to work on the bias-luminosity relation for inspiration.
For example, \cite{Beutler2012a} characterised the bias-luminosity relation for 6dFGS; this could be used to modify the cross-correlation model to account for the expected galaxy bias when considering how a given peculiar velocity responds to a particular galaxy overdensity. 
An alternative would be to modify the data directly such that the galaxy overdensity sample had a constant galaxy bias over the whole volume, as implemented by \cite{Carrick2015}.
We believe significant further research is required to implement either of these solutions for our method, which includes validation with simulations before application to data.

While modelling redshift-space distortions, we have assumed that the parallel-plane approximation holds for our data.
This approximation has been shown to break down for galaxy pairs with wide opening angles, which is common for large-area surveys at low-redshift, like 6dFGS.
We note that redshift-space distortions may be modelled without this limit, as shown by \cite{Szalay1998} and \cite{Szapudi2004}, and that such modelling was included in the original 6dFGS redshift-space distortion analysis by \cite{Beutler2012}.
To our knowledge, the cross-correlation model has not been derived without assuming the parallel-plane approximation, which would need to be done before it could be adopted self-consistently into our model.

In terms of the redshift-space distortion model, we also note that \cite{Beutler2012} use Feldman-Kaiser-Peacock weighting to improve their constraints.
Future work on the covariance model could include an investigation into implementing the weighting scheme and its effect on our constraints.

Given that \cite{Johnson2014} provided scale-dependent constraints on the growth rate of structure for peculiar velocities using a similar model formalism, we believe implementing a similar capability for the complete covariance is a natural extension of this work.
Based on the improvements we saw in the growth rate when including the cross-covariance, we  expect similar improvements in constraints of the growth rate in scale-dependent bins.
Tighter constraints would enable better tests of modified gravity models, such as those presented by \cite{Johnson2016}.

Our method is also nicely set up to look for signatures left behind by non-Gaussian perturbations present during the inflationary period of the Universe.
Such perturbations are a feature of alternative inflation models and they imprint a scale-dependent bias in the galaxy distribution. 
Consequently, any deviation from Gaussianity in the early universe modifies galaxy clustering on very large scales; the strength of the deviation is parametrized by $f_{NL}$.
While large scales are difficult to measure in low-redshift surveys (making it difficult to constrain $f_{NL}$), adding peculiar velocities and their cross-correlation with the galaxy distribution may tighten such constraints.
For example, \cite{Howlett2017} used Fisher matrix forecasts to show that the cross-correlation helps break degeneracies between $f_{NL}$ and $\beta$, which improves constraints on $f_{NL}$ by up to 40\%. 
This has already been leveraged by \cite{Ma2013}, who constrained $f_{NL}$ in the local universe using a density-velocity comparison analysis, but it would also be possible to implement it within our modelling framework.

One of the major benefits of our method is that any model where the power spectrum is proportional to the growth rate of structure could be substituted.
We could potentially use this feature to constrain the growth rate under the assumption of a specific modified gravity model (such as $f(R)$ gravity), rather than under the assumption of general relativity.
This could be done by using power spectra for modified gravity models, such as those produced by \texttt{MGCAMB} \citep{Hojjati2011}.
Such research would provide interesting insights into how growth rate of structure constraints respond to the assumed cosmological and gravitational model.

Finally, the upcoming Taipan Galaxy Survey \citep{DaCunha2017} is set to improve the redshift and peculiar velocity samples significantly, leading to better constraints of the growth rate of structure.

%The last numbered section should briefly summarise what has been done, and describe
%the final conclusions which the authors draw from their work.
\section{Summary} \label{sec:summary}
We have presented a significant advancement by adding redshift-space distortions (RSD) to the self-consistent model of the auto- and cross-covariance for the galaxy overdensity and peculiar velocity fields presented by \cite{Adams2017}.
This has allowed us to directly test whether the same growth rate drives the amplitude of peculiar velocities and RSD.
We have also performed a detailed analysis of how various model systematics affect our final growth rate constraint from 6dFGS and have provided a systematic error estimate in addition to our statistical uncertainty.

Our constraints from the complete covariance model are $f\sigma_8 = 0.384 \pm 0.052 \rm{(stat)} \pm 0.061 \rm{(sys)}$ for the growth rate of structure and $\beta = 0.289^{+0.044}_{-0.043} \rm{(stat)} \pm 0.049 \rm{(sys)}$ for the redshift-space distortion parameter.
We found that the statistical uncertainties were reduced by 64\% when compared to the galaxy overdensity auto-covariance only constraint and 50\% when compared to the logarithmic distance ratio auto-covariance only constraint.
Our current analysis provides an 18\% improvement on the statistical uncertainty in $f\sigma_8$ found by \cite{Adams2017}.
We believe this improvement is driven both by the improved model, which captures the information on the growth rate of structure encoded in the galaxy overdensity field through RSD, and by the use of a larger galaxy overdensity sample.
The fact that our systematic uncertainties are larger than our statistical uncertainties is primarily driven by the degeneracy between the growth rate and the relative effective bias between the galaxy overdensity auto-covariance and cross-covariance, which we parametrized as $\alpha_b$.
We anticipate that this could be mitigated by improving our underlying bias model to account for the fact that the cross-covariance is sensitive to a different effective bias than the galaxy overdensity auto-covariance.
We found that our constraint is consistent with the $\Lambda$CDM prediction of $f\sigma_8$ from the Planck 2015 cosmological parameters, as well as multiple analyses of galaxy overdensities and peculiar velocities from 6dFGS.
This validates our method as a new approach for constraining $f\sigma_8$ from large-scale structure and peculiar velocities.

As in \cite{Adams2017}, we see obvious improvements in the statistical uncertainty when utilising the cross-covariance compared to either auto-covariance alone, or the naive constraint that one achieves by treating the two fields as independent.
Once again, this supports the findings from the various theoretical studies on multi-tracer analyses, where accessing cross-correlations improves constraints.
Our results also motivate the application of this method to future large-scale structure and peculiar velocity surveys such as Taipan.

\section*{Acknowledgements}\label{sec:acknowledgements}
We are thankful to the referee for providing a thoughtful and constructive review of our work.
We are grateful to Paul Carter and Florian Beutler for providing the 6dFGS COLA mocks we employed in this paper.
The 6dF Galaxy Survey was made possible by contributions from many individuals towards the instrument, the survey and its science. We particularly thank Matthew Colless, Heath Jones, Will Saunders, Fred Watson, Quentin Parker, Mike Read, Lachlan Campbell, Chris Springob, Christina Magoulas, John Lucey, Jeremy Mould, and Tom Jarrett, as well as the dedicated staff of the Australian Astronomical Observatory and other members of the 6dFGS team over the years. We have used \texttt{matplotlib} \citep{Hunter2007} for the generation of scientific plots. This research was conducted by the Australian Research Council Centre of Excellence for All-sky Astrophysics (CAASTRO), through project number CE110001020. CA was supported by an Australian Government Research Training Program Scholarship.
	
%%%%%%%%%%%%%%%%%%%%%%%%%%%%%%%%%%%%%%%%%%%%%%%%%%



\bibliographystyle{mnras}
\bibliography{cadams_6dfrsdpv} % if your bibtex file is called example.bib

%\begin{thebibliography}{99}
%\end{thebibliography}

%%%%%%%%%%%%%%%%%%%%%%%%%%%%%%%%%%%%%%%%%%%%%%%%%%

%%%%%%%%%%%%%%%%% APPENDICES %%%%%%%%%%%%%%%%%%%%%

\appendix
\section{Derivation of Covariance Expressions Under RSD} \label{sec:appendix}
In this Appendix, we present the derivation of the expressions for the four covariance matrices that make up our complete model covariance, defined as 
\begin{align}
\mathbfss{C} = \begin{pmatrix}
\mathbfss{C}_{\delta \delta}  \ \mathbfss{C}_{\delta \eta} \\
\mathbfss{C}_{\eta \delta} \ \mathbfss{C}_{\eta \eta}
\end{pmatrix},
\end{align}
where $\mathbfss{C}_{\delta \delta}$ is the galaxy overdensity auto-covariance, $\mathbfss{C}_{\eta \eta}$ is the logarithmic distance ratio auto-covariance, and $\mathbfss{C}_{\delta \eta}$ and $\mathbfss{C}_{\eta \delta}$ are the cross-covariances.
This model corresponds to our chosen data vector
\begin{align}
\bm{\Delta} =  \begin{pmatrix}
	\bm{\delta} \\ \bm{\eta}
\end{pmatrix},
\end{align}
where $\bm{\Delta}$ contains the list of overdensities $\bm{\delta}_g$ and logarithmic distance ratios $\bm{\eta}$ measured from simulations or the 6-degree Field Galaxy Survey. 
In this appendix, we will present the model for peculiar velocity, which is related to the logarithmic distance ratio via the conversion factor $\xi$, defined in Eq. \ref{eq:veltoetaconversiondef}.

As we are modelling the effects of redshift-space distortions (RSD), our theoretical model for the galaxy overdensity in Fourier space is 
\begin{align}
\tilde{\delta}_g^s(\bm{k}) = [b\tilde{\delta}_m(\bm{k}) + f\mu^2\tilde{\theta}(\bm{k})]D_g(k,\mu,\sigma_g), \label{eq:app_deltag_def}
\end{align} 
and our theoretical model for the logarithmic distance ratio is 
\begin{align}
\tilde{v_p}(\bm{k}) = - iaHf\frac{\mu}{k}\tilde{\theta}(\bm{k})D_u(k,\sigma_u). \label{eq:app_eta_def}
\end{align}
Here, $\tilde{\delta}_m(\bm{k})$ is the matter overdensity field and $\tilde{\theta}(\bm{k})$ is the velocity divergence field, both in Fourier space. $b$ is the galaxy bias in real space, $f$ is the growth rate of structure, $a$ is the cosmological scale factor, and $H$ is the Hubble parameter. $D_g$ and $D_u$ are the damping functions for the RSD model, defined in terms of their respective damping parameters ($\sigma_g$ and $\sigma_u$) in Eq. \ref{eq:rsd_densitydamp_function} and \ref{eq:vkmudamp_function}.

Throughout, we use the following position conventions:
\begin{align}
\bm{x}_s &= (x_{s_x}, x_{s_y}, x_{s_z}), \ |\bm{x}_s| = x_s, \\
\bm{x}_t &= (x_{t_x}, x_{t_y}, x_{t_z}), \ |\bm{x}_t| = x_t, \\
\bm{r} &= \bm{x}_t - \bm{x}_s = (r_x, r_y , r_z), \ |\bm{r}|= r, \\
\bm{d} &= \frac{1}{2} [\bm{x}_t + \bm{x}_s] = (d_x, d_y , d_z), \ |\bm{d}|= d, \\
\hat{\bm{k}} &= (\sin \theta \cos \phi, \sin \theta \sin \phi, \cos \theta), \\
\mu &= \hat{\bm{k}} \cdot \hat{\bm{d}}.
\end{align}
We refer the reader to fig. 2 of \cite{Adams2017} for a visual representation of the configuration space vectors and angles.

Eq. \ref{eq:app_deltag_def} and \ref{eq:app_eta_def} allow us to calculate the anisotropic power spectra: 
\begin{align}
P_{gg}(k,\mu)&= b^2[\begin{aligned}[t] P_{mm}(k) &+ 2r_g\beta\mu^2P_{m\theta}(k) \\ &+\beta^2\mu^4P_{\theta\theta}(k)]D_g^2(k,\mu,\sigma_g),\end{aligned} \label{eq:app_gganisotropicps} \\
P_{gv}(k,\mu)&=\frac{iaHfb\mu}{k} [\begin{aligned}[t] r_g &P_{m\theta}(k) +\beta\mu^2 P_{\theta\theta}(k)]\\ & D_g(k,\mu,\sigma_g)D_u(k,\sigma_u), \end{aligned} \label{eq:app_gvanisotropicps} \\
P_{vg}(k,\mu)&=\frac{-iaHfb\mu}{k} [\begin{aligned}[t] r_g &P_{m\theta}(k) +\beta\mu^2 P_{\theta\theta}(k)] \\ &D_g(k,\mu,\sigma_g)D_u(k,\sigma_u), \end{aligned} \label{eq:app_vganisotropicps} \\
P_{vv}(k,\mu)&= \left(\frac{aHf\mu}{k}\right)^2 P_{\theta\theta}(k) D_u^2(k,\sigma_u), \label{eq:app_vvanisotropicps}
\end{align}
where $r_g$ is the cross-correlation coefficient discussed in the text following Eq. \ref{eq:vvanisotropicps}.

Throughout the following sections, we use a number of mathematical identities and definitions.
Given we are working with anisotropic power spectra, we make use of the multipole expansion
\begin{align}
P(k,\mu) &= \sum_{\ell=0}^{\infty} P_{\ell}(k)L_{\ell}(\mu), \label{eq:pstomultipoles}
\end{align}
where $P_{\ell}(k)$ are the multipole power spectra and $L_{\ell}(\mu)$ are the Legendre polynomials. 
Eq. \ref{eq:pstomultipoles} can then be evaluated for the required $P_{\ell}$, which is done by multiplying each side by $L_{\ell'}(\mu)$ and taking advantage of the normalisation condition for Legendre polynomials:
\begin{align}
\int_{-1}^{1} L_{\ell}(x) L_{\ell'}(x) dx = \frac{2}{2\ell'+1} \delta_{\ell\ell'},
\end{align}
such that
\begin{align}
P_{\ell}(k) &= \int_{-1}^1  \frac{2\ell +1}{2}L_{\ell}(\mu) P(k,\mu) d\mu. \label{eq:multipolepsdef}
\end{align}
The expression of the wavevector in spherical coordinates allows us utilise the plane wave expansion
\begin{align}
e^{i\bm{k}\cdot \bm{r}} = \sum_\ell i^{\ell} (2\ell+1) j_\ell(kr) L_\ell(\hat{\bm{k}}\cdot \hat{\bm{r}} ), \label{eq:planewaveexp}
\end{align}
where $j_{\ell}$ are the spherical Bessel functions and $L_{\ell}$ are the Legendre polynomials. 
It is also useful to note that any function of $\theta$ and $\phi$ may be expressed as a linear sum of spherical harmonic functions:
\begin{align}
f(\theta,\phi) = \sum_{\ell=0}^{\infty}\sum_{m=-\ell}^{\ell} f_{\ell m} Y_{\ell,m}(\theta,\phi), \label{eq:sphericalharmonicdecomp}
\end{align}
and that the coefficients can be directly calculated through
\begin{align}
f_{\ell m} = \int_{\theta=0}^{\pi} \int_{\phi=0}^{2\pi} f(\theta,\phi) Y^*_{\ell,m}(\theta,\phi) \sin(\theta) d\theta d\phi. \label{eq:sphericalharmoniccoeff}
\end{align}
We also note that our normalisation convention for spherical harmonics is such that we may define the complex conjugate of $Y_{\ell,m}(\theta,\phi)$ to be
\begin{align}
Y^*_{\ell,m}(\theta,\phi) = (-1)^mY_{\ell,-m}(\theta,\phi),
\end{align}
and that the following orthonormal condition holds:
\begin{align}
\int_{\theta=0}^{\pi} \int_{\phi=0}^{2\pi} Y_{\ell,m}(\theta,\phi)Y^*_{\ell',m'}(\theta,\phi) \sin(\theta) d\phi d\theta = \delta_{\ell,\ell'}\delta_{m,m'}, \label{eq:orthogonality}
\end{align}
where $\delta_{\ell,\ell'}$ and $\delta_{m,m'}$ are Kronecker delta functions, which evaluate to 1 if the subscripts are equal and 0 otherwise. 
Finally, we note that the spherical harmonic addition theorem is useful when working with Legendre polynomials where the argument is a dot-product of unit vectors:
\begin{align}
L_{\ell}(\hat{\bm{k}}\cdot \hat{\bm{r}}) &= \frac{4\pi}{(2\ell + 1)} \sum_{m=-\ell}^{\ell} Y_{\ell m}(\hat{\bm{k}}) Y_{\ell m}(\hat{\bm{r}})^* \nonumber \\
&= \frac{4\pi}{(2\ell + 1)} \sum_{m=-\ell}^{\ell} Y_{\ell m}(\hat{\bm{k}})^* Y_{\ell m}(\hat{\bm{r}}). \label{eq:sphericalharmonicaddition}
\end{align}

\subsection{Galaxy Overdensity Auto-Covariance} \label{subsec:rsd_dd_eqs}
Given the definition of the anisotropic power spectrum, we can write the galaxy overdensity auto-covariance as
\begin{align}
C_{\delta \delta} (\bm{x}_s, \bm{x}_t) &= \frac{1}{(2\pi)^3}\int P_{gg}(k,\mu)  e^{i\bm{k}\cdot \bm{r}} d^3\bm{k}.
\end{align}
Utilising Eq. \ref{eq:pstomultipoles} and \ref{eq:planewaveexp}:
\begin{align}
C_{\delta \delta} (\bm{x}_s, \bm{x}_t) &= \frac{1}{(2\pi)^3}\int \begin{aligned}[t] \sum_{\ell,\ell'} &P_{gg,\ell}(k) L_{\ell}(\hat{\bm{k}}\cdot\hat{\bm{d}}) (2\ell' + 1) \\ &i^{\ell'} j_{\ell'}(kr) L_{\ell'}(\hat{\bm{k}}\cdot\hat{\bm{r}}) d^3 \bm{k}. \end{aligned}
\end{align}
This can then be expanded through the spherical harmonic addition theorem (Eq. \ref{eq:sphericalharmonicaddition}):
\begin{align}
C_{\delta \delta} (\bm{x}_s, \bm{x}_t) &= \frac{1}{(2\pi)^3}\int \sum_{\ell,\ell'} \begin{aligned}[t] &\sum_{m,m'} P_{gg,\ell}(k) \frac{4\pi}{(2\ell+1)}\\ 
&Y_{\ell m}(\hat{\bm{k}}) Y_{\ell m}^*(\hat{\bm{d}}) (2\ell' + 1) i^{\ell'} j_{\ell'}(kr) \\&\frac{4\pi}{(2\ell'+1)}Y_{\ell' m'}^*(\hat{\bm{k}}) Y_{\ell' m'}(\hat{\bm{r}}) d^3 \bm{k}.\end{aligned}
\end{align}
We now break up the integral into spherical coordinates $d^3\bm{k} = k^2 \sin(\theta) d\phi d\theta dk$, noting that $\hat{\bm{k}}$ is a function of $\theta$ and $\phi$, but $\hat{\bm{d}}$ and $\hat{\bm{r}}$ are not.
This allows us to group the spherical harmonic functions into configuration-space and Fourier-space pairs:
\begin{align}
C_{\delta \delta} (\bm{x}_s, \bm{x}_t) &= \frac{1}{(2\pi)^3}\int_0^{\infty} \sum_{\ell,\ell'} \begin{aligned}[t] &\sum_{m,m'} k^2 P_{gg,\ell}(k) \frac{(4\pi)^2}{(2\ell+1)} \\&Y_{\ell m}^*(\hat{\bm{d}})Y_{\ell' m'}(\hat{\bm{r}})i^{\ell'} j_{\ell'}(kr) \\ 
& \int_0^{\pi} \int_{0}^{2\pi} Y_{\ell m}(\hat{\bm{k}})Y_{\ell' m'}^*(\hat{\bm{k}}) \\&\sin(\theta) d\phi d\theta dk.\end{aligned}
\end{align}
The angular integral corresponds to the orthonormal condition of spherical harmonics (Eq. \ref{eq:orthogonality}), producing the pair of delta functions $\delta_{\ell,\ell'}\delta_{m,m'}$ such that
\begin{align}
C_{\delta \delta} (\bm{x}_s, \bm{x}_t) &= \frac{1}{(2\pi)^3}\int_0^{\infty} \sum_{\ell} \begin{aligned}[t] &\sum_{m} k^2 P_{gg,\ell}(k) \frac{(4\pi)^2}{(2\ell+1)} \\&Y_{\ell m}^*(\hat{\bm{d}})Y_{\ell m}(\hat{\bm{r}})i^{\ell} j_{\ell}(kr) dk, \end{aligned} 
\end{align}
which can be further reduced through the spherical harmonic addition theorem to
\begin{align}
C_{\delta \delta} (\bm{x}_s, \bm{x}_t) &= \frac{1}{2\pi^2}\int_0^{\infty} \sum_{\ell} k^2 P_{gg,\ell}(k)L_{\ell}(\cos\gamma)i^{\ell} j_{\ell}(kr) dk \label{eq:ggcovariancersd}
\end{align}
where $\gamma$ is the angle between $\bm{r}$ and $\bm{d}$.

The next step is to assess which values of $\ell$ are required for the expansion, and to determine the power spectrum multipole function at the required $\ell$.
Given the form of the expansion (Eq. \ref{eq:pstomultipoles}), the required values of $\ell$ are determined by the orders of $\mu$ that appear in the anisotropic power spectrum.
For the galaxy-galaxy anisotropic power spectrum (Eq. \ref{eq:app_gganisotropicps}), the orders of $\mu$ are $0,2,4$.
Recalling the definition of the power spectrum multipoles (Eq. \ref{eq:multipolepsdef}):
\begin{align}
P_{gg,\ell}(k) &= \int_{-1}^1  \frac{2\ell +1}{2} \begin{aligned}[t] &L_{\ell}(\mu) b^2[P_{mm}(k) + 2r_g\beta\mu^2P_{m\theta}(k) + \\& \beta^2\mu^4P_{\theta\theta}(k)]D_g^2(k,\mu,\sigma_g) d\mu. \end{aligned} \label{eq:ggcovariancersdmultipoles}
\end{align}

As discussed in Section \ref{subsec:cov_model}, we break the covariance into components that can be scaled by our free parameters, which saves computing time.
The galaxy overdensity auto-covariance (given by Eq. \ref{eq:ggcovariancersd} and \ref{eq:ggcovariancersdmultipoles}) can be expressed as
\begin{align}
\mathbfss{C}_{\delta \delta} = \frac{b^2}{2\pi^2}(\mathbfss{C}_{\delta \delta, \beta^0} + 2r_g\beta\mathbfss{C}_{\delta\delta, \beta^1} + \beta^2\mathbfss{C}_{\delta \delta, \beta^2}), \label{eq:ggbetasum}
\end{align}
where each of these covariance matrices will include the sum over $\ell$ of the power spectrum multipoles as well as the integrals over $\mu$ and $k$. 
The integral over $\mu$ can be evaluated analytically, whereas the integral over $k$ is done numerically.
We obtained the analytic expressions for the various covariances through \texttt{Mathematica}.

The covariances matrices for each order of $\beta$ may then be expressed as the sum of integrand matrices for each value of $\ell$, which we label with $\mathbfss{K}$. For the $\beta^0$ term:
\begin{align}
\mathbfss{C}_{\delta \delta, \beta^0} &= \int k^2 P_{mm}(k) \bigg[\mathbfss{K}_{\delta \delta, \beta^0, \ell=0} + \begin{aligned}[t]&\mathbfss{K}_{\delta \delta, \beta^0, \ell=2} \\&+ \mathbfss{K}_{\delta \delta, \beta^0, \ell=4}\bigg] dk. \end{aligned} \label{eq:K_dd_begin}
\end{align}
For a pair of positions (described by $\gamma$ and $r$) the integrands have the following functional forms: 
\begin{align}
K_{\delta \delta, \beta^0,\ell=0} &= \frac{1}{2k\sigma_g} \sqrt{\pi}\text{Erf}(k\sigma_g)j_0(kr), \\
K_{\delta \delta, \beta^0,\ell=2} &= \begin{aligned}[t]&\frac{5}{8k^3\sigma_g^3} L_2(\cos\gamma) \bigg[ 6 e^{-k^2\sigma_g^2} k\sigma_g \\&+ (-3 + 2k^2\sigma_g^2)\sqrt{\pi}\text{Erf}(k\sigma_g) \bigg] j_2(kr), \end{aligned} \\
K_{\delta \delta, \beta^0,\ell=4} &= \begin{aligned}[t]&\frac{9}{64k^5\sigma_g^5}L_4(\cos\gamma)\bigg[  -10e^{-k^2\sigma_g^2}k\sigma_g(21 + 2k^2\sigma_g^2) \\ &+ 3(35-20k^2\sigma_g^2 + 4k^4\sigma_g^4)\sqrt{\pi}\text{Erf}(k\sigma_g) \bigg] j_4(kr),\end{aligned}
\end{align}
where $\text{Erf}(x)$ is the error function. For the $\beta^1$ term:
\begin{align}
\mathbfss{C}_{\delta \delta, \beta^1} &= \int k^2 P_{mm}(k) \bigg[\mathbfss{K}_{\delta \delta, \beta^1, \ell=0} + \begin{aligned}[t]&\mathbfss{K}_{\delta \delta, \beta^1, \ell=2} \\&+ \mathbfss{K}_{\delta \delta, \beta^1, \ell=4}\bigg] dk, \end{aligned}
\end{align}
where
\begin{align}
K_{\delta \delta, \beta^1,\ell=0} &= \begin{aligned}[t]&\frac{1}{2k^3\sigma_g^3}\bigg[ -2e^{-k^2\sigma_g^2}k\sigma_g \\&+ \sqrt{\pi}\text{Erf}(k\sigma_g)\bigg]j_0(kr), \end{aligned} \\
K_{\delta \delta, \beta^1,\ell=2} &= \frac{5}{8k^5\sigma_g^5} \begin{aligned}[t] &L_2(\cos\gamma) \bigg[ 2 e^{-k^2\sigma_g^2} k\sigma_g \\ &(9 + 4k^2\sigma_g^2) + (-9 + 2k^2\sigma_g^2)\\ &\sqrt{\pi}\text{Erf}(k\sigma_g) \bigg] j_2(kr), \end{aligned} \\
K_{\delta \delta, \beta^0,\ell=4} &= \begin{aligned}[t] &\frac{-9}{64k^7\sigma_g^7}L_4(\cos\gamma)\bigg[ 2e^{-k^2\sigma_g^2}k\sigma_g \\&(525 + 170k^2\sigma_g^2 + 32k^4\sigma_g^4) \\ &- 3(175-60k^2\sigma_g^2 + 4k^4\sigma_g^4)\\ &\sqrt{\pi}\text{Erf}(k\sigma_g) \bigg] j_4(kr).\end{aligned}
\end{align}
Finally, the $\beta^2$ term can be expressed as:
\begin{align}
\mathbfss{C}_{\delta \delta, \beta^2} &= \int k^2 P_{mm}(k) \bigg[\mathbfss{K}_{\delta \delta, \beta^2, \ell=0} +  \begin{aligned}[t]&\mathbfss{K}_{\delta \delta, \beta^2, \ell=2} \\&+ \mathbfss{K}_{\delta \delta, \beta^2, \ell=4}\bigg] dk,\end{aligned}
\end{align}
where
\begin{align}
K_{\delta \delta, \beta^2,\ell=0} &= \begin{aligned}[t]&\frac{1}{8k^5\sigma_g^5}\bigg[ -2e^{-k^2\sigma_g^2}k\sigma_g(3 + 2k^2\sigma_g^2) \\&+ 3\sqrt{\pi}\text{Erf}(k\sigma_g)\bigg]j_0(kr), \end{aligned}\\
K_{\delta \delta, \beta^2,\ell=2} &= \begin{aligned}[t]&\frac{5}{32k^7\sigma_g^7} L_2(\cos\gamma) \bigg[2 e^{-k^2\sigma_g^2} k\sigma_g \\&(45 + 24k^2\sigma_g^2 + 8k^4\sigma_g^4) \\ &+ 3(-15 + 2k^2\sigma_g^2)\sqrt{\pi}\text{Erf}(k\sigma_g) \bigg] j_2(kr), \end{aligned} \\
K_{\delta \delta, \beta^2,\ell=4} &= \begin{aligned}[t] &\frac{-9}{256k^9\sigma_g^9}L_4(\cos\gamma)\bigg[ 2e^{-k^2\sigma_g^2}k\sigma_g\\&(3675 + 1550k^2\sigma_g^2 + 416k^4\sigma_g^4 + 64k^6\sigma_g^6) \\ &- 3(1225-300k^2\sigma_g^2 + 12k^4\sigma_g^4)\\&\sqrt{\pi}\text{Erf}(k\sigma_g) \bigg] j_4(kr).\end{aligned} \label{eq:K_dd_end}
\end{align}

\subsection{Peculiar Velocity Auto-Covariance} \label{subsec:rsd_vv_eqs}
The mathematics for the peculiar velocity auto-covariance is largely the same as what we used in the previous section. 
Elements of the covariance matrix have the same form as Eq. \ref{eq:ggcovariancersd}:
\begin{align}
C_{v v} (\bm{x}_s, \bm{x}_t) &= \frac{1}{2\pi^2}\int_0^{\infty} \sum_{\ell} k^2 P_{vv,\ell}(k)L_{\ell}(\cos\gamma)i^{\ell} j_{\ell}(kr) dk \label{eq:vvcovariancersd},
\end{align}
where the multipole power spectra are given by
\begin{align}
P_{vv,\ell}(k) = \int_{-1}^1 \frac{2\ell+1}{2} L_\ell(\mu)\left(\frac{aHf\mu}{k}\right)^2P_{\theta\theta}(k) D_u^2(k,\sigma_u) d\mu. \label{eq:vvcovariancersdmultipoles}
\end{align}
For the velocity-velocity anisotropic power spectrum (Eq. \ref{eq:app_vvanisotropicps}), the orders of $\mu$ indicate that we require $\ell$ = 0, 2 for the multipole expansion.

Unlike the galaxy overdensity auto-covariance, there is only a single order of $\beta$, so we do not need to express the total covariance as a sum over orders of $\beta$, as in Eq. \ref{eq:ggbetasum}. 
Instead, we may jump straight to the expression in terms of integrand matrices $\mathbfss{K}$:
\begin{align}
\mathbfss{C}_{v v} = \frac{(aHf)^2}{2\pi^2}\int P_{\theta\theta}(k) D^2_u(k,\sigma_u) \bigg[\mathbfss{K}_{vv, \ell=0} + \mathbfss{K}_{v v, \ell=2}\bigg] dk.
\end{align}
For a pair of positions (described by $\gamma$ and $r$) the integrands have the following functional forms: 
\begin{align}
K_{v v,\ell=0} &= \frac{1}{3}j_0(kr) \label{eq:K_vv_begin} \\ 
K_{v v,\ell=2} &= -\frac{2}{3} L_2(\cos\gamma)j_2(kr). \label{eq:K_vv_end}
\end{align}

\subsection{Cross-Covariance} \label{subsec:rsd_dv_eqs}
Again, the mathematics for introducing RSD to the cross-covariance is largely the same as that used in the previous derivations.
Elements of the covariance matrix have the same form as Eq. \ref{eq:ggcovariancersd}:
\begin{align}
C_{\delta v} (\bm{x}_{\delta}, \bm{x}_v) &= C_{v \delta} (\bm{x}_{v}, \bm{x}_{\delta}) \\&= \frac{1}{2\pi^2}\int_0^{\infty} \sum_{\ell} k^2 P_{vg,\ell}(k)L_{\ell}(\cos\gamma)i^{\ell} j_{\ell}(kr) dk. \label{eq:dvcovariancersd}
\end{align}
Here, we have specified the equation in terms of the overdensity and velocity positions ($\bm{x}_{\delta}, \bm{x}_v$).
By choosing to define $\bm{r} = \bm{x}_{\delta} - \bm{x}_v$, we account for the asymmetry of the cross-covariance, which appears as a sign difference when working in terms of fixed positions ($\bm{x}_s, \bm{x}_t$). 
Given there are outstanding factors of $i$, we note that the covariance expression above will only have the correct sign if calculated using the expression for $P_{vg,\ell}$.

The multipole power spectra are given by
\begin{align}
P_{vg,\ell}(k) &= \begin{aligned}[t]\int_{-1}^1 &\frac{2\ell+1}{2} L_\ell(\mu) \frac{-iaHfb\mu}{k} [r_g P_{m\theta}(k) \\&+\beta\mu^2 P_{\theta\theta}(k)] D_u(k,\sigma_u)D_g(k,\mu,\sigma_g) d\mu.\end{aligned}
\end{align}
The orders of $\mu$ indicate that we require $\ell$ = 1, 3 for the multipole expansion.

As with the galaxy overdensity auto-covariance, there are multiple orders of $\beta$, so we break up the covariance equation similarly to Eq. \ref{eq:vgbetasum}:
\begin{align}
\mathbfss{C}_{v \delta} = \frac{aHfb}{2\pi^2}(r_g\mathbfss{C}_{v \delta, \beta^0} + \beta\mathbfss{C}_{v \delta, \beta^1}). \label{eq:vgbetasum}
\end{align}
Again, we obtain the analytic expressions for the various covariances through \texttt{Mathematica}.

The covariance matrices for each order of $\beta$ may be expressed as the sum of integrand matrices for each value of $\ell$, which we label with $\mathbfss{K}$. For the $\beta^0$ term:
\begin{align}
\mathbfss{C}_{v \delta, \beta^0} &= \int k P_{\theta m}(k) D_u(k,\sigma_u)\bigg[\mathbfss{K}_{v \delta, \beta^0, \ell=1} + \mathbfss{K}_{v \delta, \beta^0, \ell=3}\bigg] dk. \label{eq:K_dv_begin}
\end{align}
For a pair of positions (described by $\gamma$ and $r$) the integrands have the following functional forms: 
\begin{align}
K_{v \delta, \beta^0,\ell=1} &= \begin{aligned}[t] &\frac{3}{2k^3\sigma_g^3}L_1(\cos\gamma)\bigg[-2e^{-k^2\sigma_g^2/2}k\sigma_g \\ &+ \sqrt{2\pi}\text{Erf}\left(\frac{k\sigma_g}{\sqrt{2}}\right) \bigg] j_1(kr), \end{aligned} \\
K_{v \delta, \beta^0,\ell=3} &= \begin{aligned}[t] &\frac{7}{4k^5\sigma_g^5}L_3(\cos\gamma)\bigg[2e^{-k^2\sigma_g^2/2}k\sigma_g(15 + 2k^2\sigma_g^2) \\ &+ 3\sqrt{2\pi}(-5 + k^2\sigma_g^2)\text{Erf}\left(\frac{k\sigma_g}{\sqrt{2}}\right) \bigg] j_3(kr), \end{aligned}
\end{align}
where $\text{Erf}(x)$ is the error function. For the $\beta^1$ term:
\begin{align}
\mathbfss{C}_{v \delta, \beta^1} &= \int k P_{\theta m}(k) D_u(k,\sigma_u)\bigg[\mathbfss{K}_{v \delta, \beta^1, \ell=1} + \mathbfss{K}_{v \delta, \beta^1, \ell=3}\bigg] dk,
\end{align}
where
\begin{align}
K_{v \delta, \beta^1,\ell=1} &= \begin{aligned}[t] &\frac{3}{2k^5\sigma_g^5}L_1(\cos\gamma)\bigg[-2e^{-k^2\sigma_g^2/2}k\sigma_g(3 + k^2\sigma_g^2) \\ &+ 3\sqrt{2\pi}\text{Erf}\left(\frac{k\sigma_g}{\sqrt{2}}\right) \bigg] j_1(kr), \end{aligned} \\
K_{v \delta, \beta^1,\ell=3} &= \begin{aligned}[t] &\frac{7}{4k^7\sigma_g^7}L_3(\cos\gamma)\bigg[2e^{-k^2\sigma_g^2/2}k\sigma_g\\&(75 + 16k^2\sigma_g^2 + 2k^4\sigma_g^4) \\ &+ 3\sqrt{2\pi}(-25 + 3k^2\sigma_g^2)\text{Erf}\left(\frac{k\sigma_g}{\sqrt{2}}\right) \bigg] j_3(kr). \end{aligned} \label{eq:K_dv_end}
\end{align} 


%%%%%%%%%%%%%%%%%%%%%%%%%%%%%%%%%%%%%%%%%%%%%%%%%%

% Don't change these lines
\bsp    % typesetting comment
\label{lastpage}
\end{document}